\documentclass{aa}  

\usepackage{natbib}
\bibpunct{(}{)}{;}{a}{}{,} % to follow the A&A style

\usepackage{graphicx}
\usepackage[varg]{txfonts}
\usepackage{subcaption}         % necessary for continued figures, example in section 3
\usepackage{lscape}             % to rotate a single page table, example in appendix.
\usepackage{placeins}           % useful with \FloatBarrier, to keep 
\usepackage{orcidlink}	

\newcommand{\rmd}{{\rm d}}
\newcommand{\rme}{{\rm e}}
\newcommand{\rmi}{{\rm i}}

\newcommand{\four}{\mathcal{F}}
\newcommand{\fourinv}{\mathcal{F}^{-1}}
\newcommand{\del}[1]{\Delta #1}

\begin{document}

%%%%%%%%%%%%%%%%%%%%%%%%%%%%%%%%%%%%%%%%
% if you use custom commands in your title,
% ensure to check your title when submitting!
%%%%%%%%%%%%%%%%%%%%%%%%%%%%%%%%%%%%%%%%
\title{Extended vs.\ thin screen approximation in astrophysical settings}

%\subtitle{Subtitle}

%%%%%%%%%%%%%%%%%%%%%%%%%%%%%%%%%%%%%%%%
% Please separate each author with the \and command
%
% Use the \corrauth to provide the corresponding
% author address. It will be automatically inserted as 
% footnote in the PDF output.
%
% Please DO NOT include ORCIDs next to author names.
% Instead, please provide an active address for each coauthor:
% it will be automatically extracted by EDPS editorial system, 
% and co-authors will be be able to authenticate their ORCID.
%
% Only authenticated ORCIDs will be taken into account.
% ORCIDs included here will be removed.
%%%%%%%%%%%%%%%%%%%%%%%%%%%%%%%%%%%%%%%%

\author{
Manuel Adlberger\inst{}\email{manuel.adlberger@jku.at}\orcidlink{0009-0006-9075-2258}
\and Tobias Kramer\inst{}\corrauth{tobias.kramer@jku.at}\orcidlink{0000-0003-1106-3587}
}

\institute{
Institute for Theoretical Physics, Johannes Kepler University Linz, Altenberger Str.\ 69, 4040 Linz, Austria
}
\date{Received XXX, 2026}
 
\abstract
% context heading (optional)
{In the context of radio wave propagation through plasma concentrations and for gravitational lensing, the thin screen approximation often provides a suitable approximation for scattering and deflections caused by three-dimensional spatial structures.
For scattering of electromagnetic waves from an extended medium, stochastic theories provide analytical bounds on ensemble averaged quantities, such as the scintillation index.
If neither the conditions for a stochastic treatment nor the thin screen approximation are fulfilled, fully numerical methods are required.}
% aims heading (mandatory)
{For an extended structure, the applicability of the thin screen approximation needs to be verified by comparing it with the fully three-dimensional scattering solution.
We discuss various exact solutions to simple lensing geometries and provide numerical results for the propagation in the branched-flow regime, which is also of relevance in astrophysical applications. } 
% methods heading (mandatory)
{
We transfer several analytical results obtained in quantum mechanical scattering theory to astrophysical settings and complement them with  different numerical methods to quantify the deviations of thin screen theories from an extended medium results.}
% results heading (mandatory)
{We provide the analytical results for spherical waves scattered by a point source and show the emergence of branchef flow in the intermediate regime between thin screen physics and stochastic media results.}
% conclusions heading (optional), leave it empty if necessary. “Conclusions” can be used to explicit the general conclusions that can be drawn from the paper.
{}


\maketitle

\nolinenumbers

%%%%%%%%%%%%%%%%%%%%%%%%%%%%%%%%%%%%%%%%%%%%%%%%%%%%%%%%%%%%%%
\section{Introduction}
\label{sec:introduction}

Lensing of distant astronomical objects by a foreground object provides a natural magnifying glass for studying the properties of both the distant objects and the intervening ones.
For typical astrophysical applications of lensing to distant objects such as pulsars, Kirchhoff-Fresnel diffraction theory or geometric optics provide a well-established tool to calculate the intensity distributions and the time delays of signals at arriving at an observer.
In weak gravitational lensing, Fermat's principle is often used in combination with the lens equation \citep{Schneider1992, wambsganssGravitationalLensingAstronomy1998}.
In addition, the Kirchhoff-Fresnel diffraction integral can be solved numerically using Fourier methods \citep{grillo_wave_2018}, or with Picard-Lefschetz methods for a $1/r$-potential and the rational lens \citep{Feldbrugge2020, Jow2023}.
Properties to calculate include the specific intensity patterns for a lensing configuration.

If one is interested in ensemble averaged quantities across observations or scattering events, a stochastic theory of scintillations is required \citep{Rickett1984}. 
Stochastic theories provide asymptotic results for the scintillation index and insights into the general scaling behaviour with frequency.

Between the stochastic theories and the analytical results for thin screen scattering lies another parameter regime, where the three-dimensional structure of the medium affects the scattering in a characteristic way already for a single realisation.
This implies that stochastic theories are not fully applicable \citep{Heller2021,pradeepe.t._scintillometry_2025}.\\
To illustrate this, we focus on theories in which the deflection of electromagnetic waves is effectively described by a spatially varying index of refraction $n(\mathbf{r})$. 
This includes the cases of weak gravitational lenses and scattering by plasma concentrations.
We ignore the change in the polarisation direction and restrict our analysis to the scalar Helmholtz equation for a component of the electric field amplitude $u$, which needs to be solved for specific boundary conditions:
\begin{equation}\label{eq:helmholtz}
    \nabla^2 u+\frac{\omega^2}{c^2}n^2(\mathbf{r})u=0.
\end{equation}
Here, $\omega=2\pi\nu=c k$ denotes the angular frequency and $c$ the velocity of light.
This description captures both the weak field limit of gravitational lensing and the case of plasma clouds interacting with electromagnetic waves.
In the weak gravitational field approximation \citet[Eq. (2.4)]{Peters1974} the refractive index is given by
\begin{equation}
\label{eq:helmholtzgrav}
    n^2(r)=1-\frac{4\tilde{\phi}(\mathbf{r})}{c^2},
\end{equation}
where $\tilde{\phi}(\mathbf{r})$ denotes the gravitational potential.  
For a point mass $M$, $\tilde{\phi}$ becomes
\begin{equation}
    \tilde{\phi}(\mathbf{r})=-\frac{G M}{|\mathbf{r}|},
\end{equation}
where $G$ denotes the gravitational constant.

For the case of plasma lensing, the refractive index is given by
\begin{equation}
\label{eq:helmholtzplasma}
    n^2(r)=1-\frac{\omega^2_p(r)}{\omega^2},
\end{equation}
where
\begin{equation}
\label{eq:plasma_frequency}
    \omega_{\rm p}^2(\mathbf{r})=\frac{n_{\rm e}(\mathbf{r}) e^2}{\epsilon_0 m_{\rm e}}
\end{equation}
denotes the squared plasma frequency for a plasma with spatially varying electron number density $n_{\rm e}(\mathbf{r})$, elementary charge $e$, electron mass $m_{\rm e}$, and vacuum permittivity $\epsilon_0$.

To find analytic solutions of Eq.~(\ref{eq:helmholtz}), it is convenient to map the Helmholtz equation to the non-relativistic Schrödinger equation with Hamiltonian $H=-\nabla^2+V(\mathbf{r})$ and quantum mechanical (QM) parameters:
\begin{equation}
\label{eq:helmholtz_schroedinger_equivalence}
    \hbar=1,\quad m=1/2,\quad V(\mathbf{r})=\frac{\omega_{\rm p}^2(\mathbf{r})}{c^2},
    \quad E=\frac{\omega^2}{c^2}=k^2,
\end{equation}
where $m$ denotes the mass and $E$ the energy of a quantum mechanical wave that moves within the potential $V$.
Solving the Helmholtz equation is then equivalent to finding the Green's function for the stationary Schrödinger equation. 
This identification allows one to take over scattering solutions from quantum mechanics to the lensing of electromagnetic waves.
In this work, the term QM represents the full wave solution of the Helmholtz equation and does not imply inclusion of other QM effects.
Likewise, the term potential refers to the variation of the effective refractive index, either due to the gravitational potential from Eq.~(\ref{eq:helmholtzgrav}), or caused by changes in electron density within a plasma.

The manuscript is organised as follows: first, we describe lensing by a compact point mass in Sect.~\ref{sec:single_lens} using Green's functions and relate the exact solution of the Helmholtz equations to the Kirchhoff-Fresnel diffraction theory of weak gravitational lensing.
Next, we introduce a general numerical method for studying three-dimensional refractive index variations in Sect.~\ref{sec:slicing}.
As a benchmark model for the analysis of lensing in Sect.~\ref{sec:gaussian}, we discuss a Gaussian plasma lens, which is analytically solvable within the geometric optic limit \citep{Clegg1998}, as well as in the Kirchhoff-Fresnel diffraction theory 
\citep{Buckley1975}. 
Additionally, we provide a comparison of the numerical methods introduced for this example, together with a comparison between geometrical optics and wave propagation.
Finally, we discuss the application of the thin screen approach to a random medium in Sect.~\ref{sec:random_medium}, followed by a discussion of the signature of branched flow in the dynamic and secondary spectra of such a medium in Sect.~\ref{sec:secondary_spectra}.

%%%%%%%%%%%%%%%%%%%%%%%%%%%%%%%%%%%%%%%%%%%%%%%%%%%%%%%%%%%%%%
\section{Gravitational point-mass lens without and with thin screen approximation}
\label{sec:single_lens}

Lensing by a point mass illustrates the basic gravitational lensing effect and leads to features discussed by \citet{Einstein1936}, such as the Einstein ring and the increase in apparent brightness of the lensed object.
We start with a fully three-dimensional representation of the weak light-gravity interaction, as stated in Eq.~(\ref{eq:helmholtzgrav}).
Using the analogy with quantum mechanics, the solutions of Eq.~(\ref{eq:helmholtzgrav}) for a single point mass coincide with the solutions of the scattering wave functions of the hydrogen atom.

The scattering eigenfunctions can be obtained either for an incoming plane wave (corresponding to an infinite source distance) or for a finite source distance and an incoming spherical wave. 
In addition, the geometric optics limit and the semiclassical limit of the Helmholtz equation are available in analytic form.

\subsection{Plane wave solution}
\label{sub:plane_wave_solution}

For an incoming plane wave approaching a point mass, the solution of the Helmholtz equation has been adapted by \citet{Peters1974} from the Schrödinger equation.
We briefly outline this derivation based on the three-dimensional Schrödinger equation, assuming a Coulomb interaction of the form $V(\mathbf{r})=-K_c/r$.
For a plane wave $\psi_0=\exp(\rmi kz)$ travelling across the potential, the scattering solution is expressed most conveniently using cylindrical parabolic coordinates
\begin{equation}
    \tilde{\xi}\equiv r-z,\quad\text{and}\quad \tilde{\eta}\equiv r+z.
\end{equation}
The scattered wave function \citep[Eq. (22.136)]{Adam2017} is given by the Kummer confluent hypergeometric function ${}_1F_1(a,b;z)$ \citep[Eq. (13.1.2)]{Abramowitz1972}
\begin{equation}
    \psi(k,z,\tilde{\xi})=C\exp(ikz){}_1F_1(-\rmi\mu,1;ik\tilde{\xi}),
    \quad\mu=-\frac{K_c}{2k}~.
\end{equation}
This results in an amplification factor $F^\text{(plane~wave)}$ with magnitude squared
\begin{align}
\label{eq:af_adam}
    &{|F^\text{(plane~wave)}|}^2={|\psi|}^2/{|\psi_0|}^2={|C|}^2{|{}_1F_1(-\rmi\mu,1;ik\tilde{\xi})|}^2,\nonumber\\
    &{|C|}^2=\frac{2\pi\mu}{\exp(2\pi\mu)-1}.
\end{align}

For gravitational lensing, we set
\begin{equation}
\label{eq:coulomb_strenght}
    K_c=4GMk^2/c^2
\end{equation}
and obtain
\begin{equation}
    \mu=-\frac{K_c}{2k}=-\frac{4GMk}{2c^2}=-\frac{4GM\omega}{2c^3}=-\frac{\Omega}{2},
\end{equation}
where we define 
\begin{equation}
    \Omega=\frac{4GM\omega}{c^3}.
\end{equation}
The Kummer confluent hypergeometric function ${}_1F_1(a,b;z)$ also appears in the thin screen approximation, but with a different argument $z$ (see Eq.~(\ref{eq:af_fresnel-kirchhoff})).
To see this, we set the coordinate system origin to the lens's centre in the lens plane. 
Introducing the distance between the observer and the lens plane $D_{\rm L}$, the observer and the source plane $D_{\rm S}$, and the lens plane and the source plane $D_{\rm LS}$, together with the definition of the Einstein radius from Eq.~(\ref{eq:einstein_radius}), we rewrite this argument from Eq.~(\ref{eq:af_fresnel-kirchhoff}) as
\begin{align}
\label{eq:parameter_plane_wave_vs_kfd}
    \frac{\rmi}{2}\Omega y^2
    &=\frac{\rmi}{2}\frac{4GM\omega}{c^3}\left(\frac{\mathbf{Y}D_{\rm L}}{D_{\rm S}\xi_0}\right)^2\nonumber\\
    &=\frac{\rmi}{2}\frac{4GM\omega}{c^3}\left(\frac{\mathbf{Y}D_{\rm L}}{D_{\rm S}\sqrt{4GMD_{\rm L}D_{\rm LS}/(c^2D_{\rm S})}}\right)^2\nonumber\\
    &\approx\frac{\rmi}{2}\frac{\omega}{c} \left(\frac{D_{\rm L}}{D_{\rm S}\sqrt{D_{\rm L}}}\mathbf{Y}\right)^2
    =\rmi k\frac{\mathbf{X}^2}{2D_{\rm L}}
    =\rmi k\frac{X_1^2+X_2^2}{2D_{\rm L}},
\end{align}
where in the case of a plane wave $D_{\rm LS}\approx D_{\rm S}$ since $D_{\rm S}\gg D_{\rm L}$, and the variables $\mathbf{X},\mathbf{Y}$ denote the unscaled positions within the lens and the source plane, respectively.
In the case of a typical gravitational lensing system, we assume that the distance between the lens and the observer $D_{\rm L}=z$ is much larger than the range of the projected potential within the lens plane.
For $z\rightarrow\infty$, we approximate the cylindrical parabolic coordinate $\tilde{\xi}$ as
\begin{equation}
    \tilde{\xi}=(r-z)=\sqrt{X_1^2+X_2^2+z^2}-z\approx
    \frac{X_1^2+X_2^2}{2z}
\end{equation}
and obtain with this from Eq.~(\ref{eq:parameter_plane_wave_vs_kfd})
\begin{equation}
    \rmi k\tilde{\xi}\approx\frac{\rmi}{2}\Omega y^2.
\end{equation}

\subsection{Spherical wave solution}
\label{sub:green_qm}

Instead of a plane wave, we consider the case of a spherical wave emitted from a point source at position $\mathbf{r}'$ that interacts with a point mass placed at the origin of the coordinate system.
\citet{Peters1974} and \citet{Deguchi1986} also considered the case of spherical waves, corresponding to a finite distance from the source, but the authors did not give the analytic result based on the QM Green's function derived by \citet{Hostler1963}.
The QM Green's function \citep[Eq. (1.18)]{Hostler1964} for the attractive Coulomb interaction with $V(\mathbf{r})=-K_c/r$ reads
\begin{align}
    G_\text{coul}(\mathbf{r},\mathbf{r}';E)=&-\frac{2m}{\hbar^2}\frac{\Gamma(1-\rmi\tilde{\nu})}{4\pi |\mathbf{r}-\mathbf{r}'|}\nonumber\\
    &\times\left|\begin{array}{cc}
    W_{\rmi\tilde{\nu},1/2}(-\rmi \tilde{k}\alpha_+) & M_{\rmi\tilde{\nu},1/2}(-\rmi \tilde{k}\alpha_-)\\ W'_{\rmi\tilde{\nu},1/2}(-\rmi \tilde{k}\alpha_+) & M'_{\rmi\tilde{\nu},1/2}(-\rmi \tilde{k}\alpha_-)
    \end{array}\right|,
\end{align}
containing the Whittaker functions $W,\,M$ \citep[Eqs.\ (13.1.32--33)]{Abramowitz1972} and their derivatives $W',\,M'$.
In addition to setting  $\hbar=1$, $m=1/2$, the parameters of the QM solution become with $E=k^2$ and $K_c$ from Eq.~(\ref{eq:coulomb_strenght}) in terms of the gravitational constant $G$ and the mass $M$ of the lens
\begin{align}
    \alpha_\pm=|\mathbf{r}|+|\mathbf{r}'|\pm|\mathbf{r}-\mathbf{r}'|,\quad\tilde{k}=k,\quad\tilde{\nu}=\frac{2GMk}{c^2}.
\end{align}
The corresponding amplification factor becomes
\begin{equation}
\label{eq:af_qm}
    {|F^\text{(sph.~wave)}|}^2=\frac{{|G_\text{coul}(\mathbf{r},\mathbf{r}';E)|}^2}{{|G_0(\mathbf{r},\mathbf{r}';E)|}^2},
\end{equation}
where $G_0$ denotes the free Green's function \citep[Eq. (6.2.15)]{Grosche1998}
\begin{equation}
\label{eq:Green_free}
    G_0(\mathbf{r},\mathbf{r}';E)=\frac{m}{2\pi\hbar^2 |\mathbf{r}-\mathbf{r}'|}\exp\left(-\frac{\rmi\sqrt{2mE}}{\hbar}{|\mathbf{r}-\mathbf{r}'|}\right)=\frac{\rme^{-\rmi k |\mathbf{r}-\mathbf{r}'|}}{4\pi |\mathbf{r}-\mathbf{r}'|}.
\end{equation}

\subsubsection{Semiclassical approximation}
\label{sub:semiclassical_approximation}

The Whittaker functions of the exact results can be approximated using the semiclassical (SC) approach of \citet{Kanellopoulos2009b}.
Their method uses Lambert's theorem \citep{Lambert1761}, which maps all trajectories at a fixed energy between any two points in the $1/r$ potential to a straight-line motion (albeit with varying velocity along the line).
The SC Green's function is written as a sum over all classical paths from $\mathbf{r}'$ (source) to $\mathbf{r}$ (observer) weighted by the reduced classical action $S(\mathbf{r},\mathbf{r'}; E)$ and with the van~Vleck-Pauli-Morette determinant related to the classical ray density.
This prefactor diverges at the classical caustics, leading to infinities in the SC approximation.
Outside the caustics, the  SC Green's function provides a suitable approximation and establishes the link to geometrical optics.
In the case of a $1/r$ potential, two classical hyperbolae connect the source to the observer.
The SC Coulomb Green's function \citep[Eq. (B1)]{Kanellopoulos2009b} reads as follows:
\begin{align}
\label{eq:green_sc}
    G_\text{SC}(\mathbf{r},\mathbf{r}';E)=&-\frac{\rmi}{\hbar}\frac{1}{2\pi\rmi\hbar}\Big(\sqrt{|D_1|}\exp(\rmi S_1/\hbar)\nonumber\\
    &+\sqrt{|D_2|}\exp(\rmi S_2/\hbar-\rmi\pi/2)\Big),
\end{align}
where $D_1,D_2$ represents the van Vleck-Pauli-Morette determinant \citep[Eqs.\ (36,\,37)]{Kanellopoulos2009b} with
\begin{align}
    D_1&=\frac{1}{v_-v_+}\left(-\frac{m}{2|\mathbf{r}-\mathbf{r}'|}(v_-+v_+)\right)^2,\\
    D_2&=-\frac{1}{v_-v_+}\left(-\frac{m}{2|\mathbf{r}-\mathbf{r}'|}(v_--v_+)\right)^2.
\end{align}
The velocities $v_1,\,v_2$ of the hyperbolic motion depend on the distance $a$ between the hyperbola's vertex and the intersection of the two asymptotes:
\begin{align}
    &v_\pm=\sqrt{\frac{2E}{m}}\sqrt{\frac{4a-\alpha_\pm}{\alpha_\pm}},\nonumber\\
    &a=K_c/(2E),\quad K_c=4GMk^2/c^2,
    \quad E=k^2.
\end{align}
The classical actions $S_1,\,S_2$ along the corresponding two classical paths follow from \citet[Eq. (B2)]{Kanellopoulos2009b}:
\begin{align}
    S_1=&W_+- W_-,\quad S_2=W_+ + W_-,\\
    W_\pm=&\sqrt{\frac{K_cm}{a}}\Bigg(\frac{1}{2}\sqrt{(4a+\alpha_\pm)\alpha_\pm}\nonumber\\
    &+2a\ln\left(\frac{\sqrt{\alpha_\pm}+\sqrt{4a+\alpha_\pm}}{\sqrt{4a}}\right)\Bigg).
\end{align}
We obtain the corresponding amplification factor as 
\begin{equation}
\label{eq:af_sc}
    {|F^\text{(SC)}|}^2=\frac{{|G_\text{SC}(\mathbf{r},\mathbf{r}';E)|}^2}{{|G_0(\mathbf{r},\mathbf{r}';E)|}^2},
\end{equation}
In wave optics, the amplification factor is a rapidly changing function of energy $E$.
Of interest for lens observation is the averaged amplification factor \citep[Eq. (2.2)]{Alcock1997}.
\citet{Refsdal1964} introduced the thin screen approximation to calculate this amplification factor and noted that the lengthy calculation of \citet{Tikhov1938} using hyperbolic rays is not easily followed \cite[p.~23]{Refsdal1964}.

Within the SC approach, this amplification factor is readily determined by considering the upper and lower bounds of the SC Green's function.
Omitting the phase factors in the exponents, which are the root cause of the rapid oscillations, we obtain the upper ($+$) and lower ($-$) bounds of the SC Green's function from Eq.~(\ref{eq:green_sc}) as 
\begin{align}
    G_\text{SC}^{(+)}(\mathbf{r},\mathbf{r}';E)&=-\frac{\rmi}{\hbar}\frac{1}{2\pi\rmi\hbar}\left(\sqrt{|D_1|}+\sqrt{|D_2|}\right),\\ 
    G_\text{SC}^{(-)}(\mathbf{r},\mathbf{r}';E)&=-\frac{\rmi}{\hbar}\frac{1}{2\pi\rmi\hbar}\left(\sqrt{|D_1|}-\sqrt{|D_2|}\right),
\end{align}
respectively.
The averaged amplification factor is then determined by the mean value of the upper and lower bounds as
\begin{align}
    {|F_\text{mean}^\text{(SC, averaged)}|}^2&=\frac{1}{2}\frac{{|G_\text{SC}^{(+)}(\mathbf{r},\mathbf{r}';E)|}^2+{|G_\text{SC}^{(-)}(\mathbf{r},\mathbf{r}';E)|}^2}{{|G_0(\mathbf{r},\mathbf{r}';E)|}^2}\nonumber\\
    &= 4{|\mathbf{r}-\mathbf{r}'|}^2(|D_1|+|D_2|)
\end{align}
which is independent of $E$.
For a symmetric lensing system $z=D_\text{L}=-D_\text{LS}$, we obtain
\begin{equation}
\label{eq:af_sc_averaged_approximated}
    {|F_\text{approx.}^\text{(SC,averaged)}|}^2=\frac{czy^2+2\sqrt{z(c^2z^2+2GMy^2)}}{y\sqrt{z\left(8GM+c\left(czy^2+4\sqrt{z(c^2z+2GMy^2)}\right)\right)}},
\end{equation}
as a function of distance $y$ from the line of sight measured in Einstein radii.
For large distances $z$, the leading term of Eq.~(\ref{eq:af_sc_averaged_approximated}) coincides with the equation given by \citet{Einstein1936} and \citet{Refsdal1964}
\begin{equation}
\label{eq:af_einstein}
    |F^\text{(Einstein)}|^2=\frac{y^2+2}{y\sqrt{y^2+4}}.
\end{equation}

\subsubsection{Time-delay for a point mass lens}
\label{sub:time_delay}

The thin screen approach \citep[Eq. (24)]{Schneider1985} expresses the delay time as a geometric travel time along a straight line plus a contribution to the travel in the presence of actual potential.
In this case, the time delay between two images is given by \citet[Eq. (3)]{Mao1992}
\begin{equation}
\label{eq:timedelay_thinlens}
    \Delta t^\text{(thin screen)}=\frac{4GM}{c^3}\left(\frac{y}{2}\sqrt{y^2+4}+\ln\left(\frac{\sqrt{y^2+4}+y}{\sqrt{y^2+4}-y}\right)\right).
\end{equation}

Unlike in the thin screen approximation, the rays in SC quantum theory are hyperbolae, and the time delay between the two paths is given by the travel time difference of a fictitious particle with energy $\tilde{E}=mc^2/2$ as
\begin{equation}
\label{eq:timedelay_sc}
    \Delta t^\text{(SC)}= T_2-T_1=\left(\frac{\partial S_2(\tilde{E)}}{\partial \tilde{E}}-\frac{\partial S_1(\tilde{E})}{\partial \tilde{E}}\right)\Bigg|_{\tilde{E}=mc^2/2}.
\end{equation}
Considering a symmetric lensing system $z=D_\text{L}=-D_\text{LS}$, we obtain 
\begin{align}
    &\Delta t^\text{(SC)}=\frac{1}{cy^2}\left(y\sqrt{\beta}\sqrt{\alpha+y^2\beta}+\alpha\ln\left(\frac{y(\sqrt{\beta}+\sqrt{\alpha/y^2+\beta})}{\sqrt{\alpha}}\right)\right),\nonumber\\%\text{with}\quad 
    &\alpha=\frac{8GMy^2}{c^2}, %\quad\text{and}
    \quad \beta=z+\sqrt{z}\sqrt{z+\alpha}-\sqrt{z}\sqrt{4z+\alpha}.
\end{align}
For large distances $z$, the leading term of Eq.~(\ref{eq:timedelay_sc}) coincides with Eq.~(\ref{eq:timedelay_thinlens}).
The SC hyperbolae include both the geometric and the potential time delay in terms of the classical actions $S_1$ and $S_2$.

\subsection{Thin screen approximation with the Fresnel-Kirchhoff diffraction integral}
\label{sub:fresnel-kirchhoff}

The Fresnel-Kirchhoff theory is widely used in the thin screen approximation.
The lens plane is regarded as a screen from which the incoming waves are emitted, with their phases determined by projecting the three-dimensional mass density onto it.

The Fresnel-Kirchhoff diffraction integral (FKD) for the wave-field is given by a two-dimensional integral over the lens plane $\mathbf{X}$ \citep[Eq. (2.1)]{Narayan1992}
\begin{equation}
\label{eq:FKD_general}
    \psi(\mathbf{Y},z)=\frac{k}{2\pi \rmi z}\int\exp\left(\rmi\frac{k(\mathbf{X}-\mathbf{Y})^2}{2z}-\rmi\phi(\mathbf{X})\right)\rmd^2X,
\end{equation}
where $z$ denotes the position of the observer.
This function depends only on the position of the source $\mathbf{Y}$ within the source plane and the corresponding phase change $\phi(\mathbf{X})$ introduced by the screen.

In the context of gravitational lensing, these position vectors $\mathbf{X}$ and $\mathbf{Y}$ inside the corresponding planes are typically expressed in units of Einstein radii as $\mathbf{x}=\mathbf{X}/\xi_0$ and $\mathbf{y}=\mathbf{Y}/\eta_0$, with
\begin{equation}
\label{eq:einstein_radius}
    \xi_0=\sqrt{\frac{4GM}{c^2}\frac{D_LD_{LS}}{D_S}},
    \quad \eta_0=\xi_0D_S/D_L.
\end{equation}
Consequently, the FKD from Eq.~(\ref{eq:FKD_general}) can be expressed as \citep[Eq. (2)]{Feldbrugge2020}
\begin{equation}
\label{eq:Kirchhoff-Fresnel_integral}
    \psi(\mathbf{y})=\frac{\Omega}{2\pi \rmi}\int\exp(\rmi\Omega((\mathbf{x}-\mathbf{y})^2/2+\varphi(\mathbf{x})))\rmd^2x,
    \quad\Omega=\frac{4GM\omega}{c^3}. 
\end{equation}

In the case of a single gravitational point-mass lens, the corresponding lensing potential, i.e.\ the projection of the $1/r$ potential to the lens plane, reads $\varphi(\mathbf{x})=-\ln|\mathbf{x}|$.
The two-dimensional integration in Eq.~(\ref{eq:Kirchhoff-Fresnel_integral}) can be simplified to a single integration over $|\mathbf{x}|$ involving a Bessel function \citep[Eq. (2.12)]{Nakamura1999} by utilising the spherical symmetry of the lensing potential.
The remaining integral has an analytical solution \citep[Eq. (6.631.1)]{Gradstejn2009} and results in the amplification factor
\begin{equation}
\label{eq:af_fresnel-kirchhoff}
    {|F^\text{(FK)}|}^2=\frac{\pi\Omega}{1-\exp(-\pi\Omega)}{|{}_1F_1(i\Omega/2,1;i\Omega y^2/2)|}^2,
\end{equation}
involving the Kummer confluent hypergeometric function ${}_1F_1$ as described before in Eq.~(\ref{eq:af_adam}) but with different arguments due to the thin screen approximation.
In the limit of $y\rightarrow0$ and for $\Omega\gg1$, the leading term of Eq.~(\ref{eq:af_fresnel-kirchhoff}) coincides with the estimation provided by \citet{Ohanian1973} of the amplification along the line of sight of a gravitational lens using wave optics
\begin{equation}
\label{eq:max_amplification}
    |F_\text{max}|^2\approx\frac{8\pi^2GM\nu}{c^3}.
\end{equation}

\subsection{Comparison of the discussed methods}
\label{sub:comparison}

\begin{figure}
    \centering
    \includegraphics[width=\hsize]{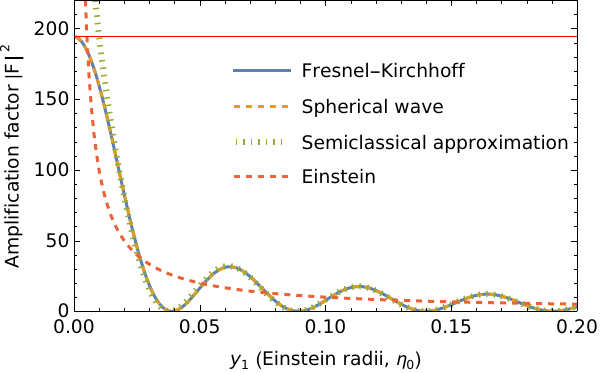}
    \caption{
    Amplification factor of a point-mass lens as a function of the source position $y_1$ from the line of sight for different methods.
    The example shows a symmetric gravitational lensing system and compares a thin screen approximation using Fresnel-Kirchhoff from Eq.~(\ref{eq:af_fresnel-kirchhoff}) with the spherical wave solution, the SC solution, and the formula given by \citet{Einstein1936} from Eqs.\ (\ref{eq:af_qm},\,\ref{eq:af_sc},\,\ref{eq:af_einstein}), respectively.
    The relative error between the spherical wave and the Fresnel-Kirchhoff solution is in the order of $10^{-13}$.
    The red line indicates the estimated amplification according to Eq.~(\ref{eq:max_amplification}).
    For large values of $y_1$, all approaches reach a limit value of 1.
    Parameters: $M=0.5~M_\odot$, $D_{\rm L}=D_{\rm LS}=1$~pc, $\nu=1$~MHz.}
    \label{fig:pointmass_qm_sc_fk}
\end{figure}

We consider a gravitational lensing system with a gravitational point-mass lens of mass $M=0.5~M_\odot$ and a symmetric arrangement of the observer, lens and source with $D_{\rm L}=D_{\rm LS}=1$~pc.
Assuming an incoming wave of frequency $\nu=1$~MHz, Fig.~\ref{fig:pointmass_qm_sc_fk} shows the amplification factor for the different methods from Eqs.\ (\ref{eq:af_qm},\,\ref{eq:af_sc},\,\ref{eq:af_einstein},\,\ref{eq:af_fresnel-kirchhoff}).
Whereas the SC solution and the formula given by \citet{Einstein1936} diverge for a perfectly rotationally symmetric lensing system along the line of sight, the limit of the spherical wave and FKD solutions approaches the estimate given in Eq.~(\ref{eq:max_amplification}).
When sufficiently far from the line of sight, all methods yield very similar results.
The FKD is a suitable approximation for distant scatterers, with a relative error of order $10^{-13}$ compared to the three-dimensional wave solution.

%%%%%%%%%%%%%%%%%%%%%%%%%%%%%%%%%%%%%%%%%%%%%%%%%%%%%%%%%%%%%%
\section{Numerical wave propagation}
\label{sec:slicing}

Solving the Helmholtz equation to obtain the energy-dependent Green's function in three spatial dimensions across large volumes requires considerable computational effort.
The energy-dependent Green's function cannot be directly obtained by diagonalising a matrix since the scattering solutions correspond to unbound solutions of the Schrödinger equation.
\citet{Kramer2008} obtain the energy-dependent Green's function from a time-dependent wave packet propagation.
This method allows one to consider either outgoing spherical waves or plane waves emitted from a source.

\subsection{Split-operator method}
\label{sub:som}

The split-operator method is a widely used algorithm for propagating a wave packet over a potential energy surface. 
A time-independent Hamiltonian $H$, consisting of a kinetic energy part $T$ and a potential energy $V$, gives rise to the time evolution operator $U(t,t')=\exp(-\rmi H(t-t'))$.
This operator propagates a wave function $\psi(t)$ from time $t'$ to $t$.
The semi-group property of the time-evolution operator allows one to divide the total time-step into $N_t$ steps
\begin{equation}
    |\psi(N_t \Delta t)\rangle = {\left[\rme^{-\rmi(\hat{T}+\hat{V})\Delta t/\hbar}\right]}^{N_t}|\psi(t=0)\rangle
\end{equation}
with $\Delta t=(t-t')/N_t$.
For a small time step, one writes the exponential as a product of non-commuting parts
\begin{equation}
    \rme^{-\rmi(\hat{T}+\hat{V})\Delta t/\hbar}
    =\rme^{-\rmi(\hat{V}/2+\hat{T}+\hat{V}/2)\Delta t/\hbar}
    \approx \rme^{-\rmi\hat{V}/2\Delta t/\hbar} \rme^{-\rmi\hat{T}\Delta t/\hbar} \rme^{-\rmi\hat{V}/2\Delta t/\hbar},
\end{equation}
which coincides with the original expression up to the third order in $\Delta t$.
For a wave function given in position space, the exponential of the potential energy operator requires a pointwise multiplication, while for the subsequent evaluation of the exponential of the kinetic energy, one must Fourier transform the wave function to momentum space. 
On an equidistant grid, the wave propagation is then a concatenation of  pointwise operations and Fourier transforms ${\cal F}$:
\begin{align}
\label{eq:som}
    \psi(\mathbf{r},N_t \Delta t) 
    \approx& ~\rme^{-\rmi\hat{V}/2\Delta t/\hbar} {\left[\rme^{-\rmi\hat{T}\Delta t/\hbar}\rme^{-\rmi\hat{V}\Delta t/\hbar}\right]}^{N_t} \rme^{\rmi\hat{V}/2\Delta t/\hbar}\psi(\mathbf{r},t=0)\nonumber\\
    =& ~\rme^{-\rmi V(\mathbf{r})/2\Delta t/\hbar} 
    {\left[\fourinv\rme^{-\rmi T(\mathbf{p})\Delta t/\hbar}\four\rme^{-\rmi V(\mathbf{r})\Delta t/\hbar}\right]}^{N_t}\nonumber\\
    &\times\rme^{\rmi V(\mathbf{r})/2\Delta t/\hbar}\psi(\mathbf{r},t=0).
\end{align}
The discrete Fourier transform implies periodic boundary conditions, leading to the wave packet re-entering at the opposite edge.
To avoid such non-physical effects, one has to choose a large enough grid and introduce an absorbing potential $V_\text{ab}$ at the grid boundaries by adding the term $\exp(-V_{\rm ab}(\mathbf{r})\Delta t/\hbar)$ after the inverse Fourier transform $\fourinv$.
During propagation, the energy-dependent Green's function is constructed by superimposing the time-dependent wave packets at the specific energy \citep{Kramer2008}
\begin{equation}
\label{eq:projected_eigenstate_wavepacket}
G(\mathbf{r},\mathbf{r}';E) \approx \Delta t\sum_{n=0}^{N_t}\rme^{\rmi E n\Delta t/\hbar} \psi(\mathbf{r},n\Delta t).
\end{equation}
A disadvantage of the method is that operators must be evaluated throughout the entire simulation domain at every time step.

\subsection{Slicing method}
\label{sub:slicing}

To reduce the amount of computation, we reduce the dimensionality of the problem by slicing the $ D$-dimensional potential into many consecutive screens of dimension $D-1$, as introduced by \citet{Fleck1976}.
Consider a two-dimensional scattering potential in the $xz$-plane and a wave function with energy $E$ and momentum $p$ that propagates along the $z$-direction.
Under the assumption of weak scattering, i.e., only small deflection angles, we approximate the time step $\Delta t$ by a spatial step $\del{z}$ using
\begin{equation}
\label{eq:forward_scat}
    \hbar k=p=mv\approx m\frac{\del{z}}{\Delta t}, \quad\ \Delta t\approx\frac{m\del{z}}{\hbar k}.
\end{equation}
Inserting Eq.~(\ref{eq:forward_scat}) into Eq.~(\ref{eq:som}) for $\hbar=1$, $m=1/2$ leads to the following solution of the Helmholtz equation
\begin{align}
\label{eq:slicing}
    \psi(x,N_z\del{z};E)
    \approx& \rme^{-\rmi V(x,N_z\del{z})/2\del{z}/(2k)} \Bigg[\prod_{n=1}^{N_z}\fourinv\rme^{-\rmi \mathbf{p}^2\del{z}/(2k)}\nonumber\\
    &\times\four\rme^{-\rmi V(x,n\del{z})\del{z}/(2k)} \Bigg]
    \rme^{\rmi V(x,z=0)/2\del{z}/(2k)}\psi(x,z=0).
\end{align}

\section{Gaussian plasma lens}
\label{sec:gaussian}

As a simple but realistic example of a plasma lens, we consider a model for a plasma cloud in the E layer of the ionosphere.
We study parameters similar to those of \citet{Wood2024} and model the plasma as a two-dimensional Gaussian potential with $\mathbf{r}=(x,z)$ using Eqs.\ (\ref{eq:plasma_frequency},\,\ref{eq:helmholtz_schroedinger_equivalence}) and a Gaussian electron density distribution $n_e(\mathbf{r})$
\begin{equation}
\label{eq:plasma_lens}
    V_\text{pl}(x,z)=V_0\exp\left(-\frac{(x^2+z^2)}{2a^2}\right),
    \quad V_0=\frac{n_{{\rm e},0} e^2}{\epsilon_0 m_{\rm e}c^2}.
\end{equation}
For the computations, we take a peak electron density of $n_{\text{e},0}=1.1284\times10^{12}~\text{m}^{-3}$ and a width of $a=1/\sqrt{2}~\text{km}$ at an altitude (along the $z$-direction) of $z_1=120~\text{km}$ above an observer on Earth.
The basic principles of gravitational lensing can be analogously applied to Gaussian plasma lenses by projecting the electron density along the line of sight \citep{Wang2025}.
To scan across a frequency band and to show the spatial structures, we focus on the dynamic spectra, defined by the intensity of the electric field across a range of frequencies $[\nu_0-\Delta_\nu/2,\nu_0+\Delta_\nu/2]$ and an observation line $[-\Delta_x/2,\Delta_x/2]$ at a fixed distance to the screen
\begin{equation}
 I(x,\nu)={|\psi(x,\nu)|}^2,
\end{equation}
 and the secondary spectra, obtained by the two-dimensional Fourier transform of the dynamic spectra:
 \begin{equation}
 \label{eq:secondary_spectrum}
{|H(c_x,c_\nu)|}^2={\left|  \int_{\nu_0-\Delta_\nu / 2}^{\nu_0+\Delta_\nu / 2} \rmd \nu\; \int_{-\Delta_x / 2}^{\Delta_x / 2} \rmd t\;\rme^{\rmi \nu c_\nu+\rmi x c_x}  I(x,\nu)\right|}^2
\end{equation}

\subsection{Fresnel-Kirchhoff diffraction theory}
\label{sub:kirchhoff-fresnel}
One widely used method is to project the two-dimensional potential onto a single one-dimensional screen and apply Fresnel-Kirchhoff diffraction theory.
For a two-dimensional Gaussian potential, the corresponding projected potential on a single screen located at $z=0$ reads
\begin{equation}
\label{eq:gaussian_proj}
    V_\text{proj}(x)=\int_\mathbb{R}V_\text{pl}(x,z)~\mathrm{d}z=a\sqrt{2\pi}V_0\exp\left(-\frac{x^2}{2a^2}\right).
\end{equation}
For a two-dimensional plane wave $\psi^\text{(2D)}(x,z\leq0)=\exp(\rmi kz)$ propagating along the $z$-axis with wave number $k$ through such a single screen, we obtain the corresponding solution for $z>0$ by solving the FKD from Eq.~(\ref{eq:FKD_general}), but only for a one-dimensional screen \citep[Eq. (6)]{Buckley1975}
\begin{equation}
\label{eq:kirchhoff-fresnel}
    \psi^\text{(2D)}(x,z)=\sqrt{\frac{k}{2\pi\rmi z}}\int_{-\infty}^\infty\exp\left(\rmi\frac{k(x-x')^2}{2z}-\rmi\phi^\text{(1D)}(x')\right)\;\rmd x',
\end{equation}
where the phase modification of the screen $\phi^\text{(1D)}$ is given in this case as
\begin{equation}
    \phi^\text{(1D)}(x')=\frac{V_\text{proj}(x')}{2k}=\phi_0\exp\left(-\frac{(x')^2}{2a^2}\right),
    \quad\phi_0=\frac{a\sqrt{2\pi}V_0}{2k}.
\end{equation}
Termwise integration of the expansion of the exponential function in Eq.~(\ref{eq:kirchhoff-fresnel}) in terms of $\phi_0$ leads to \citep[Eq. (25)]{Buckley1975}
\begin{align}
\label{eq:kirchhoff-fresnel_series}
    &\psi^\text{(2D)}(\tilde{x},\tilde{z})=\sum_{n=0}^\infty\frac{(-\rmi \phi_0)^n}{n!}(1+2n\rmi \tilde{z})^{-1/2}\exp\left(-\frac{n\tilde{x}^2}{1+2n\rmi \tilde{z}}\right),\nonumber\\
    &\tilde{x}= \frac{x}{\sqrt{2}a}, \quad \tilde{z}=\frac{z}{k2a^2}.
\end{align}

\subsection{Comparison of the different approaches}
\label{sub:som_vs_slicing}

\begin{figure}
    \includegraphics[width=\hsize]{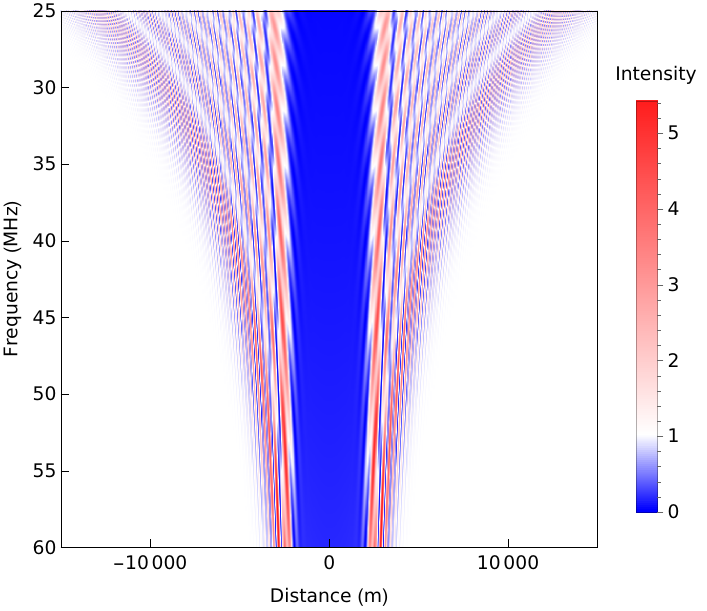}
    \hfill
    \includegraphics[width=\hsize]{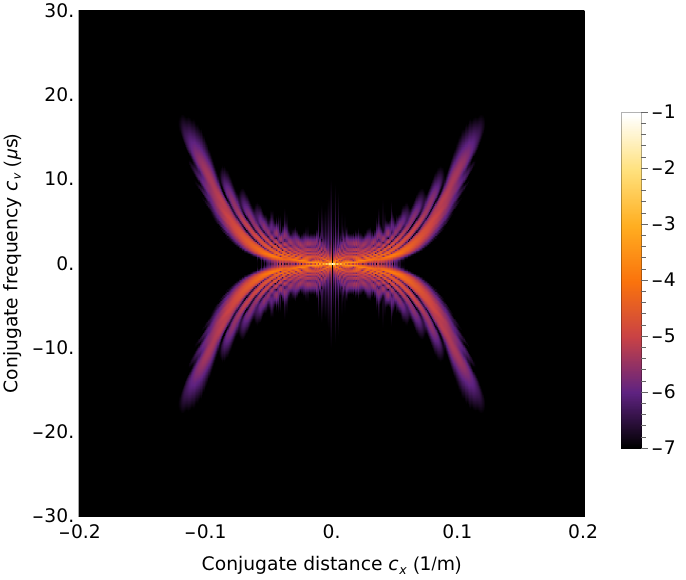}
    \caption{
    Dynamic spectrum (top) and secondary spectrum (bottom) of a Gaussian plasma lens from Eq.~(\ref{eq:plasma_lens}).
    The observed intensities at different frequencies are obtained using the slicing method from Eq.~(\ref{eq:slicing}).
    The secondary spectrum is obtained from the dynamic spectrum by Eq.~(\ref{eq:secondary_spectrum}), normalised to the maximum appearing intensity, and has a logarithmic scale.
    Parameters: $n_{\rm e,0}=1.1284\times10^{12}~\text{m}^{-3}$, $a=1/\sqrt{2}~\text{km}$, $z_1=120~\text{km}$.
    }
    \label{fig:dynamic_spectrum_elayer}
\end{figure}

For the slicing method, the two-dimensional wave function is discretised with a one-dimensional grid of 256  points.
Eq.~(\ref{eq:slicing}) is used to propagate the wave function from $z_0=-5$~km (sufficiently away from the main region of the scattering potential) through the Gaussian potential to the observer position $z_1=120$~km within a total of $N_z=40\,000$ spatial steps.
Collecting the observed intensity pattern for different radio frequencies $\nu$, as depicted in Fig.~\ref{fig:dynamic_spectrum_elayer}, one obtains a similar dynamic spectrum $I(x,\nu)$ compared to \citet{Wood2024}. 
The secondary spectrum shows curved structures, which do not appear, for instance, in a simple double-slit diffraction setup.
This indicates that the features we discuss in Sect.~\ref{sec:secondary_spectra} are related to the formation of caustics rather than interferences.

To assess the accuracy of the consecutive single-screen interaction within the slicing method, we compare it with the time-dependent propagation of a two-dimensional wave packet (split-operator method), followed by projecting an eigenstate of the corresponding energy value for $\nu=50$~MHz according to Eq.~(\ref{eq:helmholtz_schroedinger_equivalence}).

We use a Gaussian wave packet, which is infinitely extended along the $x$-direction, such that it represents a plane wave, with an initial kick $k$ along the $z$-direction as
\begin{equation}
\label{eq:gaussian_wavepacket}
    \psi(x,z,t=0)=\exp\left(-\frac{(z-z_0)^2}{2w_0^2}\right)\exp(\rmi k(z-z_0)),
\end{equation}
where $w_0=5$~m is the initial width, which we chose such that the wave packet does not disperse too fast.
We discretise this initial wave packet on a two-dimensional grid of $256\times65\,536$ points in the $x$- and $ z$-directions, respectively.
This is necessary to cover the entire propagation distance without encountering issues within the momentum grid.
Consequently, compared to the slicing method, we have a significantly larger number of grid points and a greater computational effort.
To test the split-operator method of Eq.~(\ref{eq:som}), we propagate this wave packet on the grid for $N_t=70\,000$ time steps $\Delta t=1$, project the corresponding eigenstate of the energy $E$ using Eq.~(\ref{eq:projected_eigenstate_wavepacket}), and obtain the values of the energy-dependent wave function along the points on the grid that belong to the observer position $z_1$.

\begin{figure}
    \centering
    \includegraphics[width=\hsize]{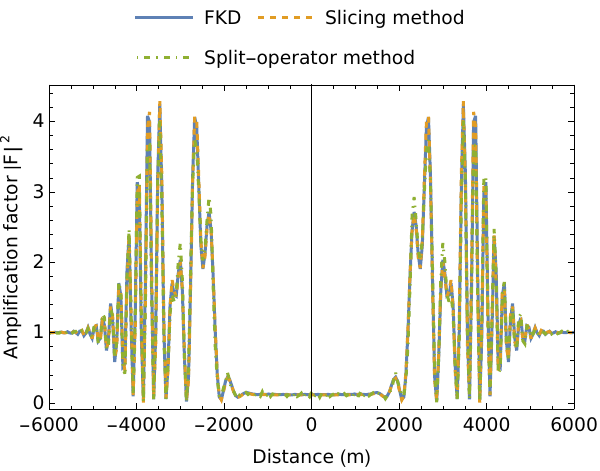}
    \caption{
    Amplification factor of a wave function scattered by a Gaussian plasma lens from Eq.~(\ref{eq:plasma_lens}) as a function of the distance from the line of sight for different methods.
    This includes the FKD (projected single 1D screen), the slicing method (many consecutive 1D screens), and the split operator method (full 2D wave packet propagation) from Eqs.\ (\ref{eq:kirchhoff-fresnel_series},\,\eqref{eq:slicing},\,\eqref{eq:projected_eigenstate_wavepacket}), respectively.
    Parameters: $n_{\rm e,0}=1.1284\times10^{12}~\text{m}^{-3}$, $a=1/\sqrt{2}~\text{km}$, $\nu=50~\text{MHz}$, $z_1=120~\text{km}$.
    }
    \label{fig:slicing_comparison}
\end{figure}

Comparing all the different methods described above, as depicted in Fig.~\ref{fig:slicing_comparison}, we observe similar solutions for the given scattering problem, but with significantly different computational efforts.
Consequently, the slicing method provides a fast approximation but only considers forward scattering.
Since we focus only on very weak scattering potentials in Sect.~\ref{sec:random_medium}, we use the slicing method to study the wave propagation through a more complex random potential landscape later on.

\subsection{Comparison between classical trajectories and the wave solution}
\label{sub:classical_vs_qm_trajectories}

We discuss some differences between classical trajectories (geometrical optics) and the regions of constructive interference within the wave solution.
Given a scattering potential $V$, we compute rays of classical trajectories for a particle of mass $m$ and initial velocity $\mathbf{v}$ by solving Newton's equation of motion numerically and compare the density of classical rays to the wave solution using the slicing method, as described above.

\begin{figure}
    \centering
    \includegraphics[width=\hsize]{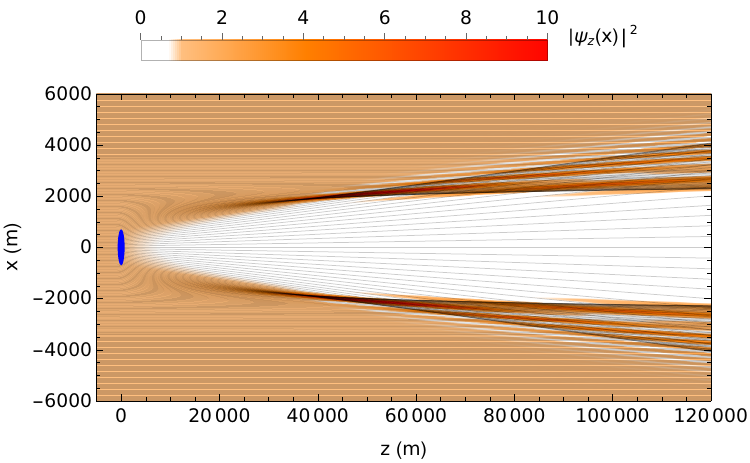}
    \caption{
    Scattering of a bundle of classical trajectories (black lines) by a single Gaussian plasma lens (blue disk with radius $a$) from Eq.~(\ref{eq:plasma_lens}) compared to the wave solution (background intensity).
    The wave solution is obtained using the slicing method from Eq.~(\ref{eq:slicing}).
    Parameters: $n_{\rm e,0}=1.1284\times10^{12}~\text{m}^{-3}$, $a=1/\sqrt{2}~\text{km}$, $\nu=50~\text{MHz}$.
    }
    \label{fig:classical_vs_wave_solution}
\end{figure}

As the first example, depicted in Fig.~\ref{fig:classical_vs_wave_solution}, we consider a single two-dimensional Gaussian plasma lens within the $xz$-plane from Eq.~(\ref{eq:plasma_lens}).
The corresponding bundle of classical rays and the plane wave have an initial wave number $k$ in the $z$-direction that corresponds to a radio frequency of $\nu=50$~MHz.

We compare the wave pattern with the intensity (ray density) distribution of a bundle of classical rays.
In the case of a one-dimensional (projected) Gaussian potential from Eq.~(\ref{eq:gaussian_proj}), analytical solutions are also available for the classical trajectories, as well as the corresponding intensity distribution observed at a distance $z$ from the screen \citep[Eq. (18)]{Clegg1998}.
\citet{Clegg1998} used a Gaussian potential of the form $\exp(-x^2/\Tilde{a}^2)$, which translates  to $\Tilde{a}=\sqrt{2}a$ in Eq.~(\ref{eq:gaussian_proj}).
The parameter $\alpha$ \citep[Eq. (14)]{Clegg1998} in the analytic solution becomes 
\begin{equation}
    \alpha=\frac{2z(V_\text{proj}(0)/(2k))}{k\Tilde{a}^2}=\frac{zV_\text{proj}(0)}{2(k a)^2}.
\end{equation}

\begin{figure}
    \centering
    \includegraphics[width=\hsize]{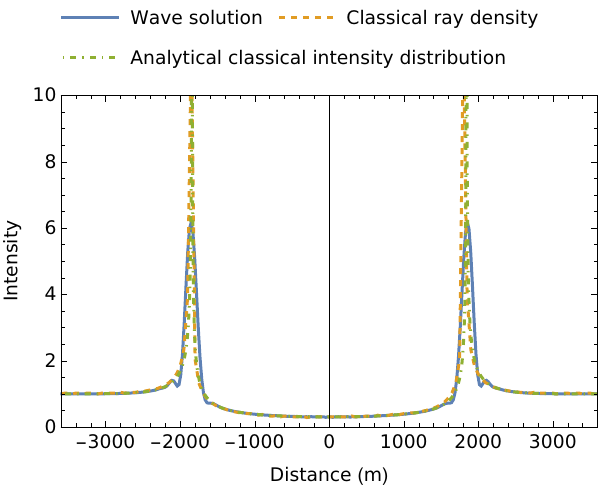}
    \hfill
    \includegraphics[width=\hsize]{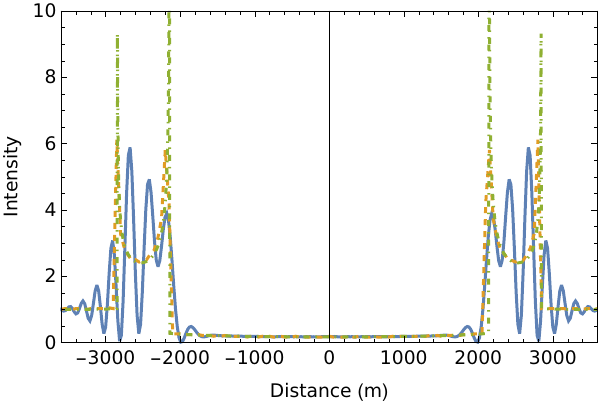}
    \caption{
    Intensity distribution as a function of the distance from the line of sight of the wave solution compared to a bundle of classical trajectories for a single Gaussian plasma lens from Eq.~(\ref{eq:plasma_lens}) at different propagation distances $z_1=35$~km (top) and $z_1=75$~km (bottom).
    The wave solution is obtained using the slicing method from Eq.~(\ref{eq:slicing}). 
    The classical intensity distribution was obtained from the analytic solution within the geometric optic limit by \citet{Clegg1998}.
    Parameters: $n_{\rm e,0}=1.1284\times10^{12}~\text{m}^{-3}$, $a=1/\sqrt{2}~\text{km}$, $\nu=50~\text{MHz}$.
    }
    \label{fig:qm_vs_classical_intensity}
\end{figure}

Figure~\ref{fig:qm_vs_classical_intensity} shows a comparison of the classical ray density and the wave intensities at different observation distances from the centre of the plasma lens.
The classical result exhibits a sudden transition between defocusing, focusing, and non-focusing regions, separated by the outer and inner caustics (regions of diverging ray density), as shown in the right panel, while the wave solution shows a smooth transition between the individual regions without diverging peaks due to the finite position-momentum uncertainty.
Note that in the left panel, both caustics nearly overlap because of the small distance to the lens.
A uniform expansion around the caustics, including third-order terms, has been performed by \citet{cordes_lensing_2017}.

%%%%%%%%%%%%%%%%%%%%%%%%%%%%%%%%%%%%%%%%%%%%%%%%%%%%%%%%%%%%%%
\section{Thin screen approach in a random medium}
\label{sec:random_medium}

Next, we apply the numerical methods developed to spatially varying plasma densities represented by a random potential $V_{\rm rp}(\mathbf{r})$.
We consider a Gaussian potential given in terms of a Fourier series and thus repeats periodically (see App.~\ref{app:gaussian_random_field} for details).
Kolmogorov phase screens have been studied by \citet{Buckley1975} or \citet{Hamidouche2007}.
We consider a varying electron density $n_{\rm e}(\mathbf{r})$ that represents a corresponding potential according to Eqs.\ (\ref{eq:plasma_frequency},\,\ref{eq:helmholtz_schroedinger_equivalence}), with a mean value of zero, 
a given variance $\sigma_{\rm V}^2$, and an underlying Gaussian correlation in two dimensions $\mathbf{r}=(x,z)$.
The corresponding autocorrelation function $\rho$ has the following form
 \begin{equation}
 \label{eq:gaussian_correlation}
     \rho(\mathbf{r})=\int_{\mathbb{R}^2}\mathrm{d^2r'V_{\rm rp}(\mathbf{r}')V_{\rm rp}(\mathbf{r}'+\mathbf{r})}=\sigma_{\rm V}^2\exp(-\mathbf{r}^2/l_{\rm c}^2)
 \end{equation}
where $l_{\rm c}$ defines the correlation length.
For weak scattering, the resulting intensity patterns lead to branched flow; see \citet{Heller2021} for a review and \citet{Pidwerbetsky1988} and \citet{Cordes1986} for early examples of co-propagating rays.

\subsection{Branched flow in wave optics}
\label{sub:qm_branched_flow}

\begin{figure}
    \centering
    \includegraphics[width=0.98\hsize]{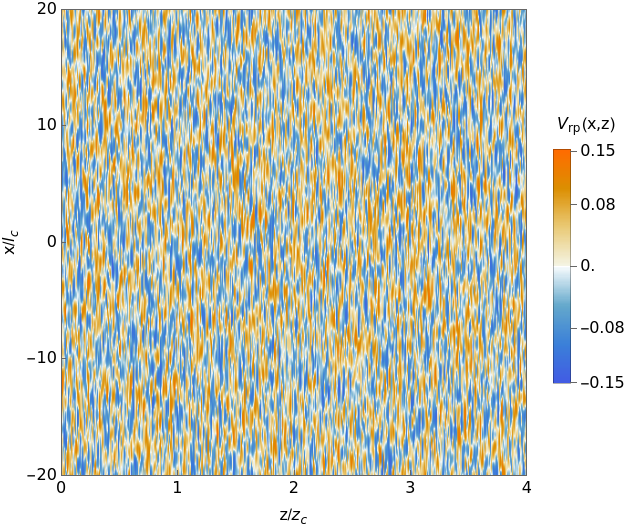}
    \hfill
    \includegraphics[width=\hsize]{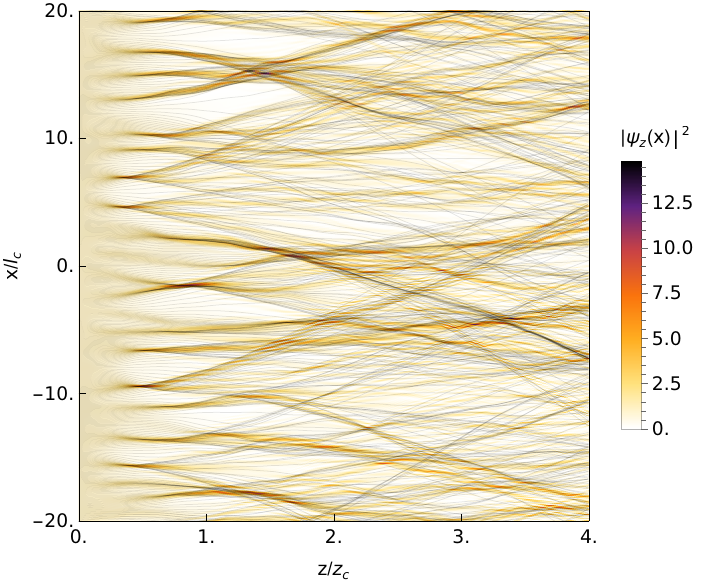}
    \caption{
    Branched flow through a random Gaussian potential.
    Top: Visualization of a random Gaussian potential $V_\text{rp}(x,z)$ from Eq.~(\ref{eq:continuous_random_potential}) along the $x$- and $z$-direction in units of $l_{\rm c}$ and $z_{\rm c}$ from Eq.~(\ref{eq:zc}), respectively.
    Bottom: Branched flow of classical trajectories (black lines) compared to the wave solution (background) while propagating through this random Gaussian potential using the slicing method from Eq.~(\ref{eq:slicing}).
    Parameters: $k=2\pi$~m${}^{-1}$, $l_{\rm c}=100$~m, $\sigma_{\rm V}=0.04$~m${}^{-2}$.
    }
    \label{fig:classical_vs_qm_branched_flow}
\end{figure}

We consider an incoming plane wave $\psi_{z=z_0}(x)$ entering at $z_0=0$.
Figure~\ref{fig:classical_vs_qm_branched_flow} shows the resulting intensity distribution for a typical realisation of a random Gaussian potential.
The intensity is obtained using the slicing method to propagate the wave function on a discrete one-dimensional grid $\{x_i\}$ of size $N_x=512$ along the $z$-axis with a step size of $\Delta z=l_{\rm c}/2$.
Superimposed on the wave pattern are the geometric optic solutions, which show bundles of almost co-propagating trajectories, the signature of branched flow \citep{Heller2021}.
Branched flow occurs for random, but correlated disorder potentials.

In an ensemble average across many random potential realisations with the same correlation function, the scintillation index $s_I^2$ is a commonly used measure to describe intensity fluctuations.
It is defined by the relative variance of the intensity distribution $I(x)=|\psi_z(x)|^2$ \citep[Eq. (2)]{Garnier2025}
\begin{equation}
\label{eq:scintillation_index}
    s_I^2(z)= \frac{\langle \sigma_I^2\rangle}{\langle \mu_I\rangle^2},
\end{equation}
where $\langle\cdot\rangle$ denotes the average over an ensemble of realisations of the random potential, $\mu_I$ the mean intensity over the grid, and $\sigma_I^2$ the unbiased estimate of variance as
\begin{equation}
    \sigma_I^2=\frac{1}{N_x-1}\sum_{i=1}^{N_x}(I(x_i)-\mu_I)^2.
\end{equation}

In the geometric optics approximation, the scintillation index in a random Gaussian potential shows a primary peak within a strongly fluctuating region due to caustics and then slowly decays to a saturated fluctuation and a limit value of 1 for very large propagation distances \citep{Metzger2014}.

\begin{figure}
    \centering
    \includegraphics[width=\hsize]{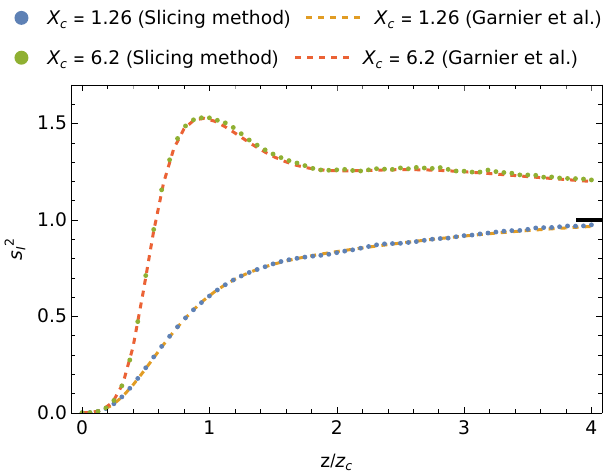}
    \caption{
    Scintillation index as a function of the propagation distance $z$ in units of $z_{\rm c}$ from Eq. \eqref{eq:zc} for an ensemble average compared to the stochastic theory given by \citet{Garnier2025}.
    The ensemble contained 1\,000 realisations of the random Gaussian potential from Eq.~(\ref{eq:continuous_random_potential}) for a given $X_{\rm c}$ from Eq. \eqref{eq:xc} and used the slicing method from Eq.~(\ref{eq:slicing}). 
    For both values of $X_{\rm c}$, the scintillation index reaches a limit value of 1 (black line) at very large propagation distances.
    Parameters: $k=2\pi$~m${}^{-1}$, $l_{\rm c}=100$~m, $\sigma_{\rm V}=0.04$~m${}^{-2}$ ($X_{\rm c}=1.26$), $\sigma_{\rm V}=0.055$~m${}^{-2}$ ($X_{\rm c}=6.2$).
    }
    \label{fig:slicing_vs_garnier}
\end{figure}

The wave optics result through a random medium obtained by \citet{Garnier2025} reveals an initial peak, followed by a smaller secondary peak.
These peaks within the scintillation index appear only if the dimensionless parameter
\begin{equation}
\label{eq:xc}
    X_{\rm c}=2^{-1/3}l_{\rm c}k(\sigma_{\rm V}/k^2)^{2/3}
\end{equation}
lies above a threshold value, which is between one and three for a Gaussian correlated random medium.
Figure~\ref{fig:slicing_vs_garnier} shows a comparison of the ensemble-averaged computations with the stochastic theory given by \citet{Garnier2025}.
We find an initial peak around $z=z_{\rm c}$, with 
\begin{equation}
\label{eq:zc}
    z_{\rm c}=2^{1/3}l_{\rm c}(\sigma_{\rm V}/k^2)^{-2/3},
\end{equation}
which is related to the $2/3$ exponent found in the classical branching length \citep{Spetzler2001, Kaplan2002}, followed by a smaller secondary peak around $z=2.5~z_{\rm c}$ before the scintillation index converges to a limit value of 1 \citep{Garnier2025}.
For lower values of $X_{\rm c}$, the peak is absent, and the scintillation index increases monotonously to 1. 

\subsection{Projected random medium}
\label{sub:projected_potential}

Next, we discuss the thin screen approximation to the random potential, resulting in a single-screen interaction.
We solve the Fresnel-Kirchhoff diffraction integral from Eq.~(\ref{eq:kirchhoff-fresnel}) for the random potential landscape $V_{\rm rp}(x,z)$ projected along the propagation direction
\begin{equation}
\label{eq:projected_potential}
    V_\text{proj}(x)=\int_\mathbb{R}V_{\rm rp}(x,z)\mathrm{d}{z}.
\end{equation}
With this projected potential, we obtain a single screen with a phase deviation of
\begin{equation}
    \phi(x)=\frac{V_\text{proj}(x)}{2k}.
\end{equation}
According to \citet[Eq. (48)]{Mercier1962}, the limit value of the scintillation index of a single phase-changing screen at very large distances reads
\begin{equation}
\label{eq:mercier_limit}
    \lim_{z\rightarrow\infty}s_I^2(z)=1-\exp(-2\phi_{\rm RMS}^2),
\end{equation}
which depends only on the root mean square (RMS) phase deviation
\begin{equation}
\label{eq:rms_phase_deviation}
    \phi_{\rm RMS}^2=\frac{1}{N_x}\sum_{i=1}^{N_x}\phi^2(x_i).
\end{equation}
For very small distances from the screen to the observer, as discussed by \citet{Buckley1975} and \citet{Salpeter1967}, the scintillation index of such a random screen follows a quadratic behaviour
\begin{equation}
\label{eq:si_approx}
    s_I^2(z)\approx\left(\frac{z~\phi_{\rm RMS}}{k~l_{\rm c}^2L_4^2}\right)^2,
\end{equation}
with $L_4=12^{-1/4}$ in the case of a Gaussian correlation.

\begin{figure}
    \centering
    \includegraphics[width=\hsize]{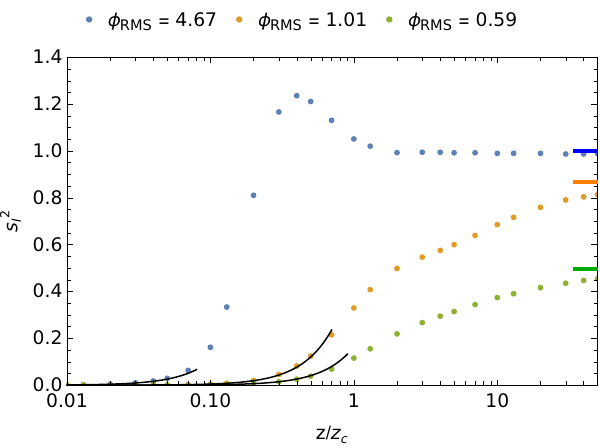}
    \caption{
    Scintillation index as a function of the propagation distance $z$ in units of $z_{\rm c}$ from Eq. \eqref{eq:zc} for an ensemble average computed in the projected single-screen approximation for different values of $\phi_{\rm RMS}$ from Eq. \eqref{eq:rms_phase_deviation}.
    The ensemble contained 5\,000 realisations of the random Gaussian potential from Eq.~(\ref{eq:continuous_random_potential}) projected on a single screen.
    The black lines indicate the approximation at small propagation distances from Eq.~(\ref{eq:si_approx}) and the coloured lines indicate the corresponding Mercier limits according to Eq.~(\ref{eq:mercier_limit}).
    From top to bottom, the Mercier limits are 1, 0.87, and 0.5.}
    \label{fig:scintillation_index_projected}
\end{figure}

Under the thin screen approximation, as shown in Fig.~\ref{fig:scintillation_index_projected}, the limit value of the scintillation index depends on the projected phase variation of the random potential and produces scintillation indices smaller than 1 if $\phi_\text{RMS}$ lies below a threshold value, which is between one and three \citep{Buckley1975}.
This behaviour is not observable in propagation through the extended two-dimensional potential landscape.

\subsection{Finite areas of a random medium}
\label{sub:localized_randpot}

The limits of an extended medium filling the space from the scattering region to the observer, on the one hand, and the approximation of a thin screen with negligible thickness, on the other, present extreme cases of more commonly encountered scenarios.

\begin{figure}
    \centering
    \includegraphics[width=\hsize]{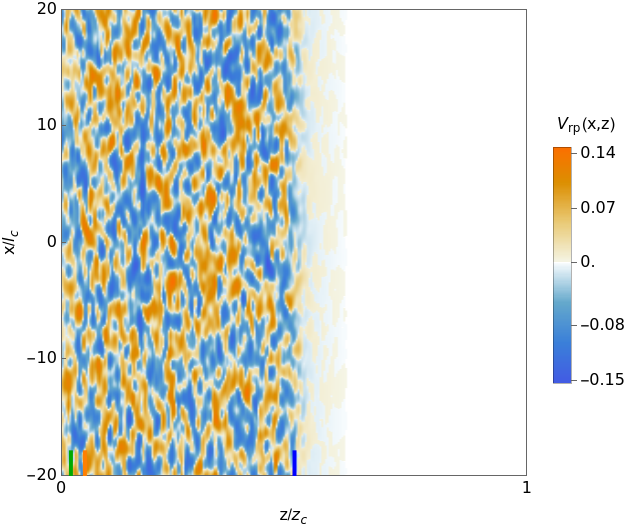}
    \hfill
    \includegraphics[width=\hsize]{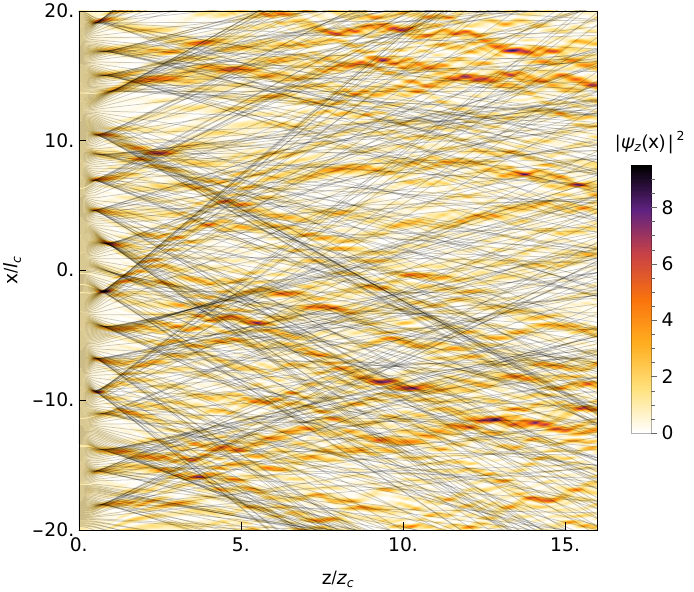}
    \caption{
    Branched flow through a spatially restricted random Gaussian potential.
    Top: Visualisation of a spatially restricted random Gaussian potential $V_\text{loc}(x,z)$ from Eq.~(\ref{eq:localized_random_potential}) of width $\Delta L=0.5~z_{\rm c}$ along the $x$- and $z$-direction in units of $l_{\rm c}$ and $z_{\rm c}$ from Eq.~(\ref{eq:zc}), respectively.
    The coloured lines indicate the corresponding upper bound $L_1$ of the random Gaussian potential along the $z$-direction (lower bound $L_0=0$ for all cases) used within Fig.~\ref{fig:scintillation_index_localized_potential}.
    From left to right, the values of $L_1$ in units of $z_{\rm c}$ are 0.02, 0.05, and 0.5.
    Bottom: Branched flow of the classical trajectories (black lines) compared to the wave solution (background) after leaving the spatially restricted random Gaussian potential.
    Parameters: $k=2\pi$~m${}^{-1}$, $l_{\rm c}=100$~m, $\sigma_{\rm V}=0.04$~m${}^{-2}$.
    }
    \label{fig:localized_potential}
\end{figure}

We next consider a spatially restricted random Gaussian potential $V_\text{loc}(x,z)$ along the propagation direction with $z\in[L_0, L_1]$ (see App.~\ref{app:gaussian_random_field}).
Figure~\ref{fig:localized_potential} shows the propagation of a plane wave that enters and leaves the scattering region computed with the slicing method of Eq.~(\ref{eq:slicing}).

\begin{figure}
    \centering
    \includegraphics[width=\hsize]{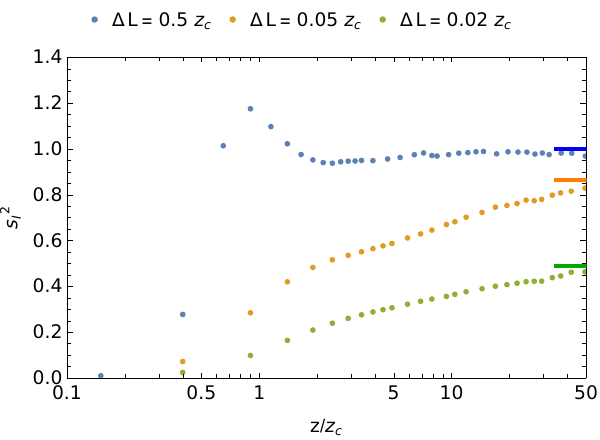}
    \caption{
    Scintillation index as a function of the propagation distance $z$ in units of $z_{\rm c}$ for an ensemble average when passing through regions of different size $\Delta L=L_1-L_0$ as indicated in Fig.~\ref{fig:localized_potential}.
    The ensemble contained 1\,000 realisations of the corresponding spatially restricted random Gaussian potential from Eq.~(\ref{eq:localized_random_potential}).
    The coloured lines indicate the corresponding Mercier limits according to Eq.~(\ref{eq:mercier_limit}).
    From top to bottom, the Mercier limits are 1, 0.86, and 0.49.
    Parameters: $k=2\pi$~m${}^{-1}$, $l_{\rm c}=100$~m, $\sigma_{\rm V}=0.04$~m${}^{-2}$.}
    \label{fig:scintillation_index_localized_potential}
\end{figure}

Figure~\ref{fig:scintillation_index_localized_potential} shows a comparison of the scintillation index for different widths of the localised region $\Delta L=L_1-L_0$ and the expectations of the scintillation index based on the projected density using the Mercier expression.
For a small width of the spatially restricted random Gaussian potential, the scintillation index approaches a limit below 1, while for increasing thickness, the limiting value 1 is assumed.

%%%%%%%%%%%%%%%%%%%%%%%%%%%%%%%%%%%%%%%%%%%%%%%%%%%%%%%%%%%%%%
\section{Signature of branched flow in dynamic and secondary spectra}
\label{sec:secondary_spectra}

\begin{figure*}
    \centering
    \includegraphics[width=0.5\linewidth]{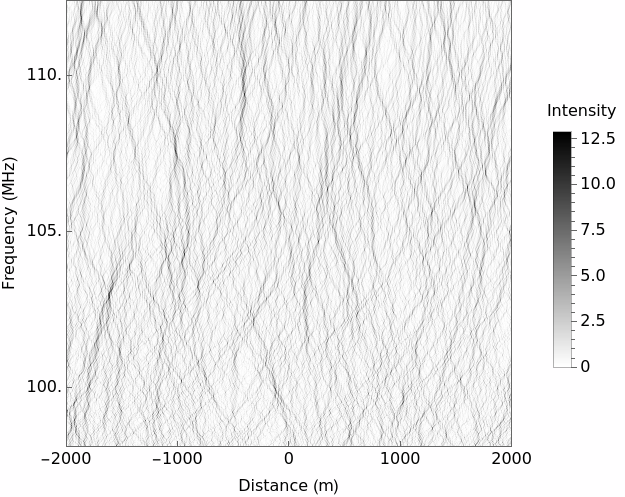}
    \hfill
    \includegraphics[width=0.48\linewidth]{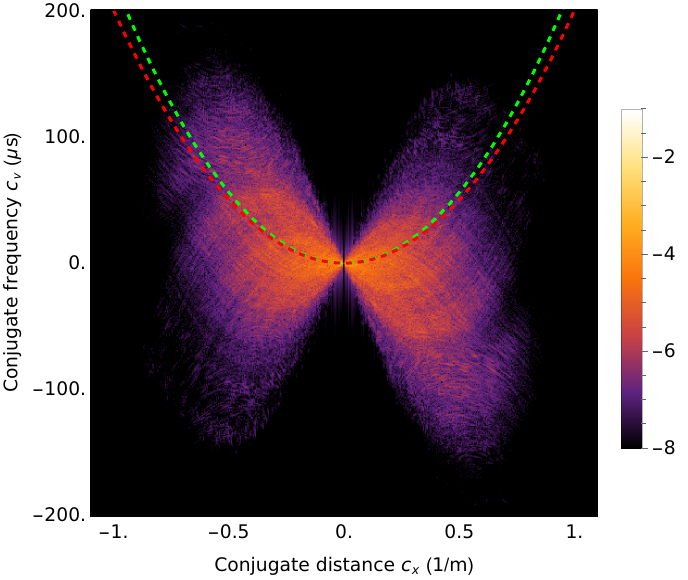}\\
    \includegraphics[width=0.5\linewidth]{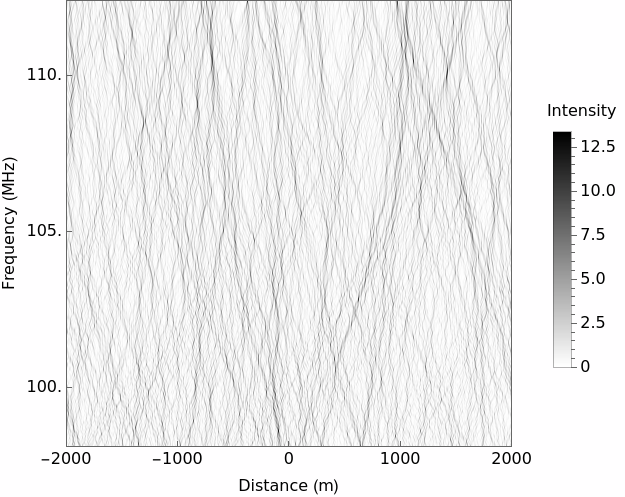}
    \hfill
    \includegraphics[width=0.48\linewidth]{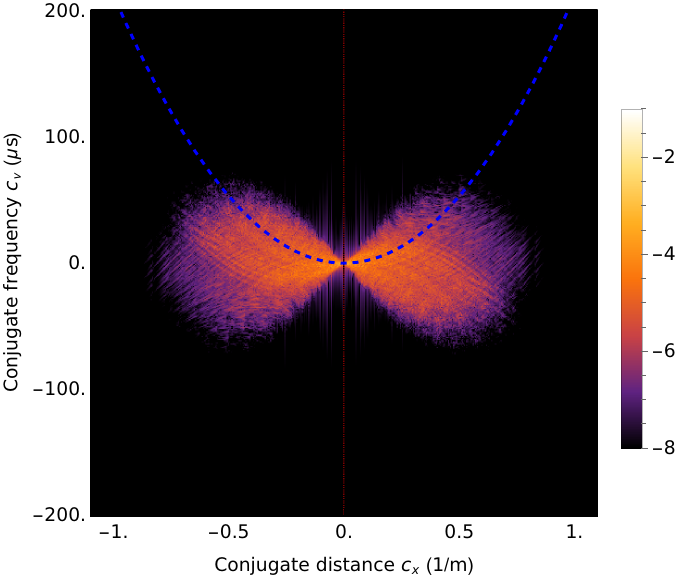}\\
    \caption{
    Dynamic spectra (left) and secondary spectra (right) of an extended random Gaussian potential compared to a projected one.
    The dynamic spectra are recorded at a distance $z_\text{obs}=10~z_{\rm c}$ (in units of $z_{\rm c}$ from Eq.~(\ref{eq:zc}) for $k=2\pi$) obtained by the slicing method from Eq.~(\ref{eq:slicing}) as a function of the distance $x$ from the line of sight and the frequency $\nu$ of the observed signal.
    The corresponding secondary spectra from Eq.~(\ref{eq:secondary_spectrum}) are normalised to the maximum appearing intensity and have a logarithmic scale.
    Top row: Results of a spatially restricted random Gaussian potential from Eq.~(\ref{eq:localized_random_potential}) of width $\Delta L=L_1-L_0=1~z_{\rm c}$.
    The secondary spectrum reveals a parabolic structure from Eq.~(\ref{eq:parabolic_arc}) (dashed green line for $z_\text{screen}=L_0=0$; dashed red line for $z_\text{screen}=L_1$).
    Bottom row: Results of the corresponding projected random potential located at $z_\text{screen}=(L_1+L_0)/2$ with a parabolic structure (dashed blue line).
    Parameters: $k_0=3\pi/4$~m${}^{-1}$, $l_{\rm c}=100$~m, $\sigma_{\rm V}=0.04$~m${}^{-1}$.
    }
    \label{fig:secondary_spectrum_randpot}
\end{figure*}

The ensemble-averaged quantities allow one to determine the scintillation index but provide no direct insight into the effects of a specific realisation of the random potential and individual branches of the intensity flow.
The branched flow structure impacts the dynamic spectra $I(x,\nu)$, which record the intensity at a fixed distance $z_\text{obs}$ from the disorder potential as a function of frequency $\nu=ck/(2\pi)$ and lateral position $x$.
The two-sided Fourier transform of the dynamic spectra yields the secondary spectra $H(c_x,c_\nu)$ as a function of the conjugate distance $c_x$ and the conjugate frequency $c_\nu$ according to Eq.~(\ref{eq:secondary_spectrum}),
which displays, for disorder concentrated at a specific distance to the observer, a parabolic scintillation arc.
\citet{Stinebring2001} observed the first scintillation arcs of pulsar signals scattered by plasma fluctuations in the interstellar medium.
Long-term monitoring of such interstellar scintillation at low frequencies can provide valuable insights into the characteristics of the interstellar medium \citep{Wu2022, Sprenger2026}.\\
Scintillation arcs have also been observed as a result of plasma variations in the Earth's ionosphere \citep{Fallows2014}.
The curvature of these parabolic arcs \citep[Eq. (8)]{Stinebring2001} depends only on the observed wave number $k_0$, the distance between the screen and the observer $D_{\rm s}$, and the distance between the source and the observer $D$.
For an incoming plane wave, set $D\gg D_{\rm s}=(z_\text{obs}-z_\text{screen})$ and one obtains
\begin{equation}
\label{eq:parabolic_arc}
    c_\nu=\frac{z_\text{obs}-z_\text{screen}}{4\pi c}\left(\frac{2\pi}{k_0}\right)^2c_x^2.
\end{equation}
Taking into account a spatially restricted random Gaussian potential from Sect.~\ref{sub:localized_randpot} with $\Delta L=z_c$ (see Appendix~\ref{app:exemplary_cases} for further examples), we use the slicing method of Eq.~(\ref{eq:slicing}) to propagate wave functions starting at $z=0$ to the observer position $z_\text{obs}$ with initial wave numbers sampled over $N_k=1024$ equidistant points between $k=k_0\pm d_k$, where we set $k_0=3\pi/4$~m${}^{-1}$ and $d_k=0.3$~m${}^{-1}$.
The top row of Fig.~\ref{fig:secondary_spectrum_randpot} shows the dynamic and secondary spectrum far beyond the random Gaussian potential, where we observe a parabolic pattern in the latter.
The secondary spectrum shows an asymmetric intensity distribution, i.e., one arm of the parabola is more extended than the other.
This behaviour is also observable for smaller widths $\Delta L$ of a spatially restricted random Gaussian potential.
If we consider the corresponding projected random potential using Eq.~(\ref{eq:projected_potential}) and solve the Fresnel-Kirchhoff diffraction integral from Eq.~(\ref{eq:kirchhoff-fresnel}), the resulting secondary spectra from the bottom row of Fig.~\ref{fig:secondary_spectrum_randpot} show a generally more symmetric pattern along both arms of the parabola.
The preserved correlation between consecutive single-screen interactions when using the slicing method appears to be an important factor in the observation of asymmetric patterns in the secondary spectrum.

%%%%%%%%%%%%%%%%%%%%%%%%%%%%%%%%%%%%%%%%%%%%%%%%%%%%%%%%%%%%%%
\section{Conclusion}
\label{sec:conclusion}
We have compared full three-dimensional wave-optics treatments with thin-screen models. 
For a point-mass lens, the thin screen Fresnel–Kirchhoff diffraction closely matches the full three-dimensional solution, 
with a relative error of order $10^{-13}$, establishing FKD as an excellent far-field approximation. 
We introduced and validated a numerically efficient slicing method for forward scattering that agrees well with a full split-operator computation while greatly reducing computational cost, enabling the subsequent random-media analysis.

In Gaussian-correlated random media, we computed the ensemble-averaged scintillation index and confirmed the initial peak near 
$z=z_{\rm c}$, a smaller secondary peak near $2.5~z_{\rm c}$, and convergence to the limiting value $1$ at large propagation distances, consistent with stochastic theory. 
Projecting an extended medium to a single thin phase screen changes the asymptotics: the Mercier limit depends on the RMS phase variations and can yield $s_I^2 < 1$, a behaviour not observed for extended propagation.
Bridging these extremes, we analysed finite-thickness random regions and found that thin regions saturate below 1, while increasing thickness drives the limit toward 1.

As a benchmark, we compared FKD, slicing, split-operator, and classical rays for a Gaussian plasma lens.
Our results delineate when thin-screen models suffice and when extended-medium effects are essential, and provide validated numerical tools to test stochastic theories across astrophysical scattering scenarios.

A direct application of the slicing method is the calculation of dynamic and secondary spectra of pulsar signals scattered by the interstellar medium. 
The relative motion of the pulsar, the interstellar medium, and Earth leads to Doppler-type spectra that show parabolic scintillation arcs with asymmetries caused by correlated disorder potentials.

%%%%%%%%%%%%%%%%%%%%%%%%%%%%%%%%%%%%%%%%%%%%%%%%%%%%%%%%%%%%%%
\section*{Data availability}

The data underlying this article will be shared on reasonable request to the authors.

\begin{acknowledgements}
This work benefited from the NVIDIA Academic Hardware Grant ``Simulating branched flow with tensor processing units'', PI Kramer (2022). 
Additional simulations were performed on the MUSICA supercomputer at Johannes-Kepler University Linz, Austria.
The authors acknowledge helpful discussions with E.J.~Heller.
\end{acknowledgements}

%%%%%%%%%%%%%%%%%%%%%%%%%%%%%%%%%%%%%%%%%%%%%%%%%%%%%%%%%%%%%%
% WARNING
% Please note that we have included the references below in
% order to compile the document, but we ask you to:
%
% - use BibTeX with the regular commands:
%   \bibliographystyle{aa} % style aa.bst
%   \bibliography{Yourfile} % your references Yourfile.bib
% - join the .bib files when you upload your source files
%%%%%%%%%%%%%%%%%%%%%%%%%%%%%%%%%%%%%%%%%%%%%%%%%%%%%%%%%%%%%%

\bibliographystyle{aa}

%%%%%%%%%%%%%%%%%%%%%%%%%%%%%%%%%%%%%%%%%%%%%%%%%%%%%%%%%%%%%%%
% Appendices must be placed after   \end{thebibliography}
% They will be placed automatically on a new page.
%%%%%%%%%%%%%%%%%%%%%%%%%%%%%%%%%%%%%%%%%%%%%%%%%%%%%%%%%%%%%%%
\begin{appendix}

\onecolumn

\section{Random Gaussian potential}
\label{app:gaussian_random_field}

To obtain a random potential with the desired Gaussian autocorrelation function \eqref{eq:gaussian_correlation} for a two-dimensional potential, we convolute white noise $W$ with an appropriate Gaussian kernel.
This convolution can be efficiently performed in momentum space, where the convolution reduces to a simple multiplication, using a fast Fourier transform (FFT) on a discrete grid as
\begin{equation}
    \tilde{V}(\{\mathbf{k}_i\})=W(\{\mathbf{k}_i\})\sqrt{P(\{\mathbf{k}_i\}})
\end{equation}
where we use the power spectral density $P(\mathbf{k})$, i.e., the Fourier transform of $\rho(\mathbf{r})$
\begin{equation}
\label{eq:gaussian_psd}
    P(\mathbf{k})=\pi l_{\rm c}^2\sigma_{\rm V}^2\exp(-\mathbf{k}^2l_{\rm c}^2/4)
\end{equation}
as the convolution kernel.
To obtain the required statistics, we rescale the potential after the inverse Fourier transform to the required variance $\sigma_{\rm V}^2$ and shift it so that we obtain the zero mean.
\begin{equation}
\label{eq:discrete_random_potential}
    V_\text{rp}(\{\mathbf{r}_i\})=\sigma_{\rm V}\frac{V(\{\mathbf{r}_i\})-\mu}{\sqrt{\sigma^2}},
\end{equation}
with $\mu$ and $\sigma^2$ being the mean and variance, respectively, of the discrete random Gaussian field $V(\{\mathbf{r}_i\})$.
The Fourier-based construction of the potential leads to periodic boundary conditions.

From this discrete potential, we construct a continuous function using a cosine Fourier series, which is differentiable and integrable. 
In the case of two dimensions, the Fourier series becomes:
\begin{equation}
\label{eq:continuous_random_potential}
    V_\text{rp}(x,z)=\sum_{m=0}^{N_x}\sum_{n=0}^{N_z}\beta_m\beta_nA_{mn}\cos\left(\frac{m\pi (x-L_x/2)}{L_x}\right)\cos\left(\frac{n\pi z}{L_z}\right)
\end{equation}
with $A_{mn}$ being the corresponding Fourier coefficients obtained from \eqref{eq:discrete_random_potential}, $\beta_0=1/2$, and $\beta_j=1$ for $j\geq1$.

To compute the random potential landscape within $(x,z)\in[-20,20]l_{\rm c}\times[0,4]z_{\rm c}$, as used for the computations in this paper, we use a discrete grid with $N_x=512$ and $N_z=1\,024$. 
For a given $\sigma_{\rm V}$ and $l_{\rm c}$, we set $L_x=40~l_{\rm c}$ and $L_z=4~z_{\rm c}$ in units of $z_{\rm c}$ from Eq.~(\ref{eq:zc}).

To localise the random field along the propagation direction to a region $z\in[L_0, L_1]$, we use an envelope function as
\begin{equation}
\label{eq:localized_random_potential}
    V_\text{loc}(x,z)=V_\text{rp}(x,z)\cosh^{-2}\left[\left(\frac{z-(L_1+L_0)/2}{(L_1-L_0)/2}\right)^8\right].
\end{equation}
This ensures a smooth transition into and out of the region of the random potential and reduces sudden changes in momentum.

\section{Exemplary cases for branched flow in extended media}
\label{app:exemplary_cases}

This section contains additional examples from Sect.~\ref{sec:secondary_spectra} of spatially restricted random Gaussian potentials with different widths $\Delta L$, illustrating the asymmetry in the intensity distribution when using the slicing method, compared with the increased symmetry obtained by using the corresponding projected potential and solving the Fresnel-Kirchhoff diffraction integral.

\begin{figure*}
    \centering
    \includegraphics[width=0.5\linewidth]{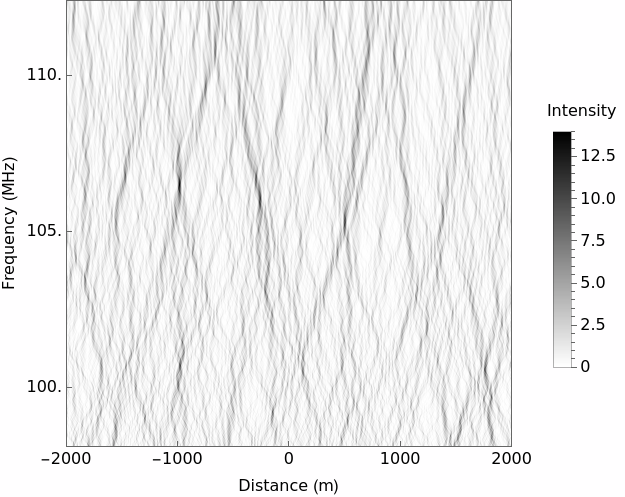}
    \hfill
    \includegraphics[width=0.48\linewidth]{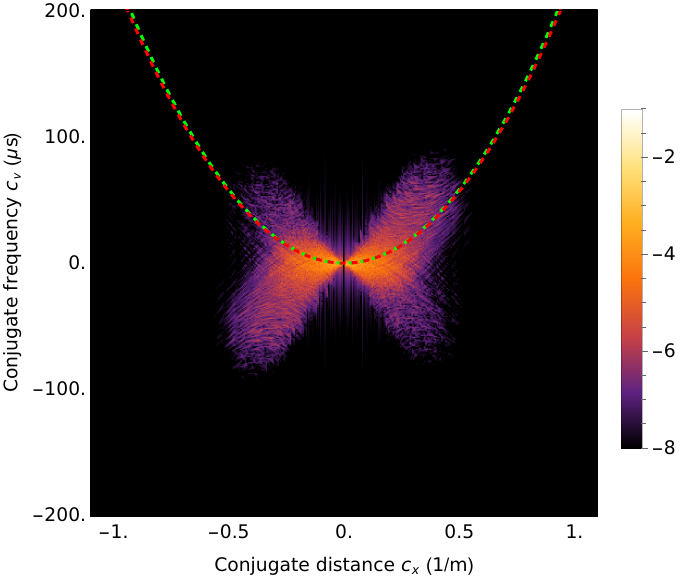}\\
    \includegraphics[width=0.5\linewidth]{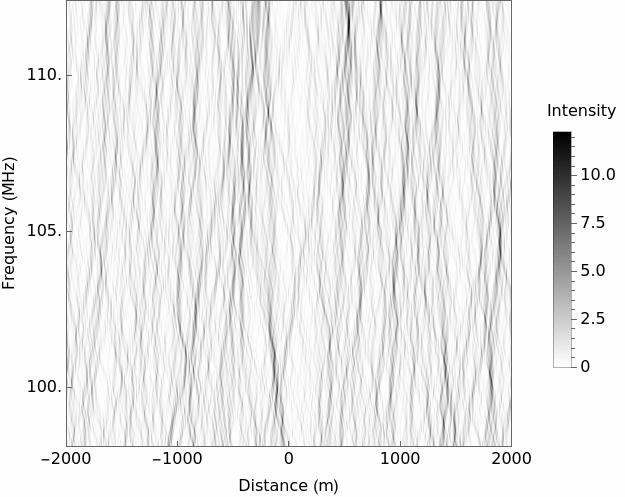}
    \hfill
    \includegraphics[width=0.48\linewidth]{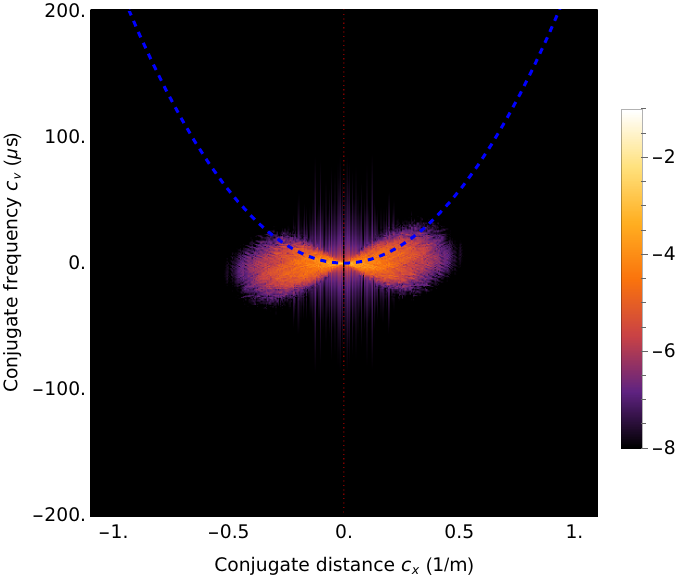}\\
    \caption{
    Dynamic spectra (left) and secondary spectra (right) of an extended random Gaussian potential compared to a projected one.
    The dynamic spectra are recorded at a distance $z_\text{obs}=10~z_{\rm c}$ (in units of $z_{\rm c}$ from Eq.~(\ref{eq:zc}) for $k=2\pi$~m${}^{-1}$) obtained by the slicing method from Eq.~(\ref{eq:slicing}) as a function of the distance $x$ from the line of sight and the frequency $\nu$ of the observed signal.
    The corresponding secondary spectra from Eq.~(\ref{eq:secondary_spectrum}) are normalised to the maximum appearing intensity and have a logarithmic scale.
    Top row: Results of a spatially restricted random Gaussian potential from Eq.~(\ref{eq:localized_random_potential}) of width $\Delta L=L_1-L_0=0.25~z_{\rm c}$.
    The secondary spectrum reveals a parabolic structure from Eq.~(\ref{eq:parabolic_arc}) (dashed green line for $z_\text{screen}=L_0=0$; dashed red line for $z_\text{screen}=L_1$).
    Bottom row: Results of the corresponding projected random potential located at $z_\text{screen}=(L_1+L_0)/2$ with a parabolic structure (dashed blue line).
    Parameters: $k_0=3\pi/4$~m${}^{-1}$, $l_{\rm c}=100$~m, $\sigma_{\rm V}=0.04$~m${}^{-2}$.
    }
\end{figure*}

\begin{figure*}
    \centering
    \includegraphics[width=0.5\linewidth]{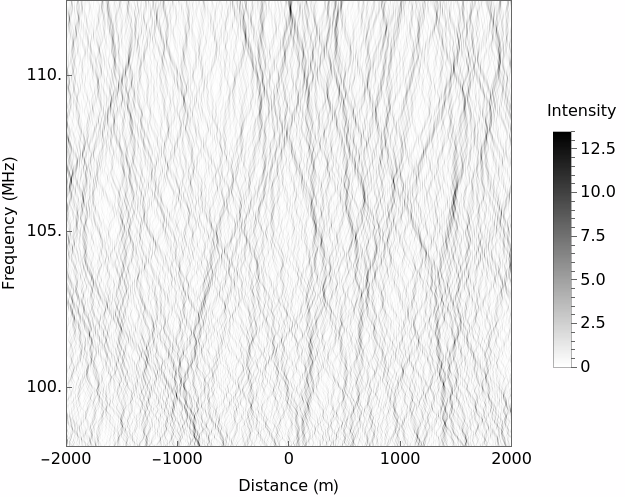}
    \hfill
    \includegraphics[width=0.48\linewidth]{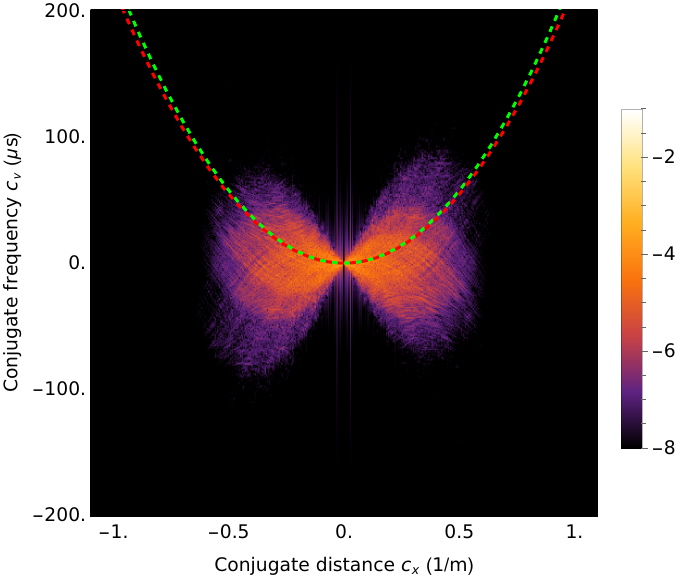}\\
    \includegraphics[width=0.5\linewidth]{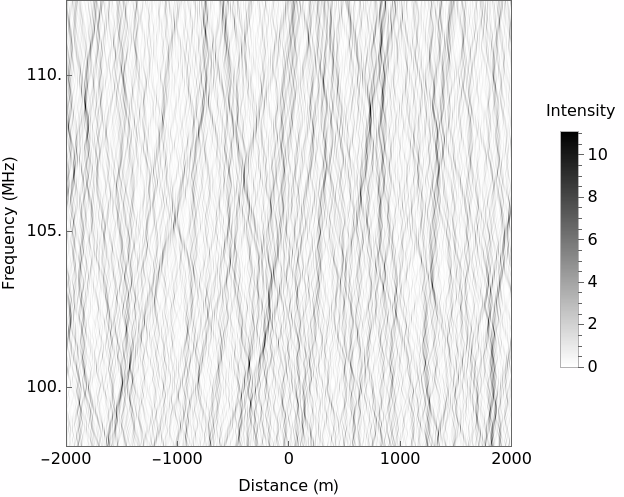}
    \hfill
    \includegraphics[width=0.48\linewidth]{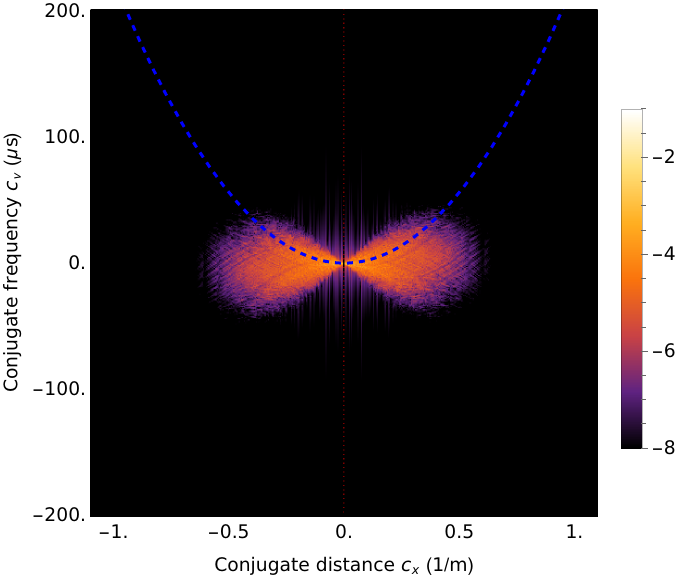}\\
    \caption{
    Dynamic spectra (left) and secondary spectra (right) of an extended random Gaussian potential compared to a projected one.
    The dynamic spectra are recorded at a distance $z_\text{obs}=10~z_{\rm c}$ (in units of $z_{\rm c}$ from Eq.~(\ref{eq:zc}) for $k=2\pi$~m${}^{-1}$) obtained by the slicing method from Eq.~(\ref{eq:slicing}) as a function of the distance $x$ from the line of sight and the frequency $\nu$ of the observed signal.
    The corresponding secondary spectra from Eq.~(\ref{eq:secondary_spectrum}) are normalised to the maximum appearing intensity and have a logarithmic scale.
    Top row: Results of a spatially restricted random Gaussian potential from Eq.~(\ref{eq:localized_random_potential}) of width $\Delta L=L_1-L_0=0.5~z_{\rm c}$.
    The secondary spectrum reveals a parabolic structure from Eq.~(\ref{eq:parabolic_arc}) (dashed green line for $z_\text{screen}=L_0=0$; dashed red line for $z_\text{screen}=L_1$).
    Bottom row: Results of the corresponding projected random potential located at $z_\text{screen}=(L_1+L_0)/2$ with a parabolic structure (dashed blue line).
    Parameters: $k_0=3\pi/4$~m${}^{-1}$, $l_{\rm c}=100$~m, $\sigma_{\rm V}=0.04$~m${}^{-2}$.
    }
\end{figure*}

\begin{figure*}
    \centering
    \includegraphics[width=0.5\linewidth]{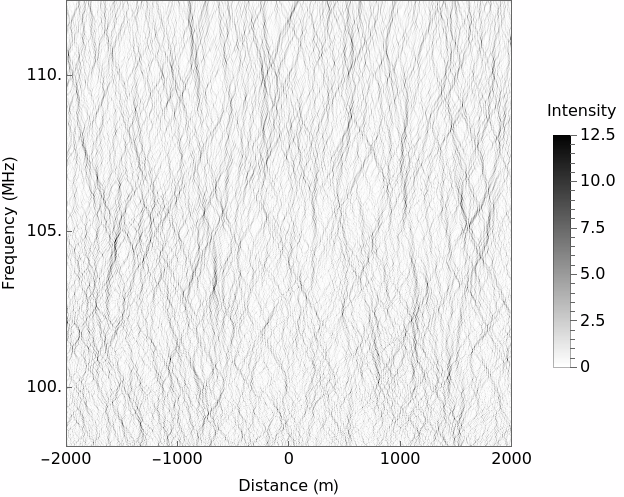}
    \hfill
    \includegraphics[width=0.48\linewidth]{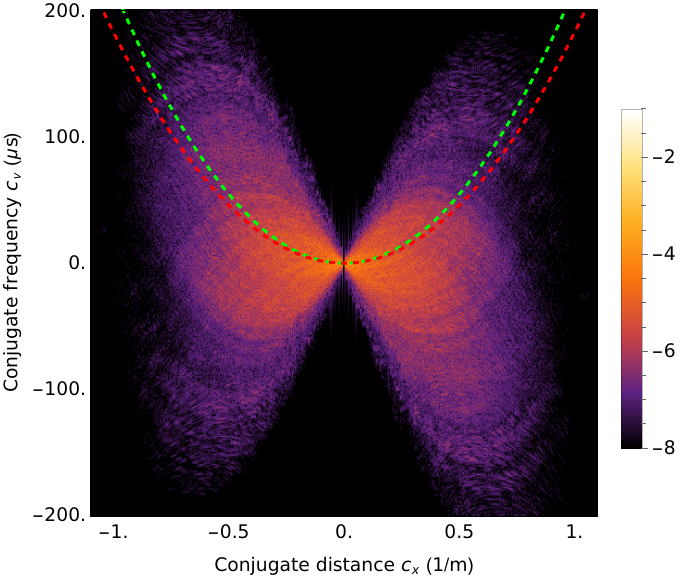}\\
    \includegraphics[width=0.5\linewidth]{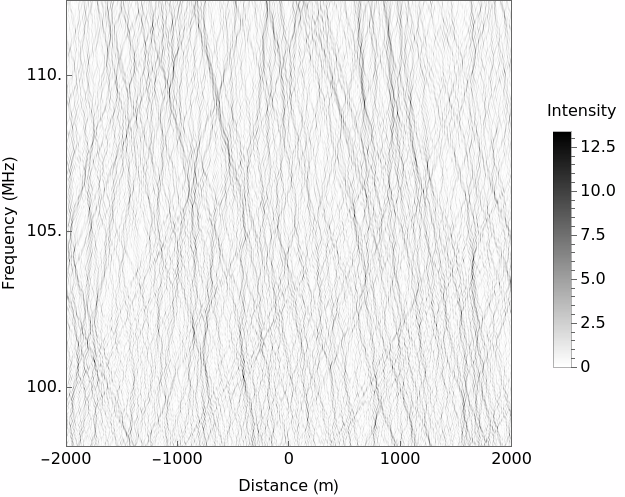}
    \hfill
    \includegraphics[width=0.48\linewidth]{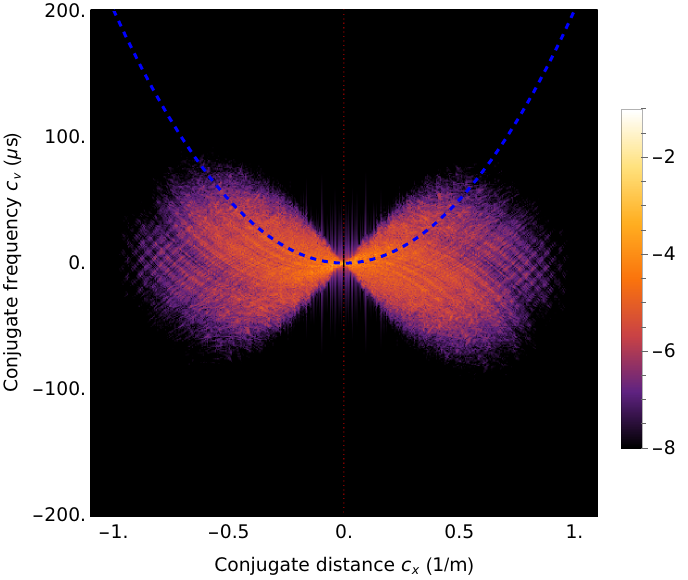}\\
    \caption{
    Dynamic spectra (left) and secondary spectra (right) of an extended random Gaussian potential compared to a projected one.
    The dynamic spectra are recorded at a distance $z_\text{obs}=10~z_{\rm c}$ (in units of $z_{\rm c}$ from Eq.~(\ref{eq:zc}) for $k=2\pi$~m${}^{-1}$) obtained by the slicing method from Eq.~(\ref{eq:slicing}) as a function of the distance $x$ from the line of sight and the frequency $\nu$ of the observed signal.
    The corresponding secondary spectra from Eq.~(\ref{eq:secondary_spectrum}) are normalised to the maximum appearing intensity and have a logarithmic scale.
    Top row: Results of a spatially restricted random Gaussian potential from Eq.~(\ref{eq:localized_random_potential}) of width $\Delta L=L_1-L_0=1.5~z_{\rm c}$.
    The secondary spectrum reveals a parabolic structure from Eq.~(\ref{eq:parabolic_arc}) (dashed green line for $z_\text{screen}=L_0=0$; dashed red line for $z_\text{screen}=L_1$).
    Bottom row: Results of the corresponding projected random potential located at $z_\text{screen}=(L_1+L_0)/2$ with a parabolic structure (dashed blue line).
    Parameters: $k_0=3\pi/4$~m${}^{-1}$, $l_{\rm c}=100$~m, $\sigma_{\rm V}=0.04$~m${}^{-2}$.
    }
\end{figure*}

\end{appendix}

\end{document}